\documentclass[aps,prl,twocolumn,showpacs,
10pt]{revtex4-2}
\usepackage{subfiles}
\usepackage{amssymb,amsmath}
\usepackage{bm, graphicx, amsmath}
\usepackage{amsthm}
\usepackage{array}
\usepackage{bbm}
\usepackage{xcolor}
\usepackage[section]{placeins}
\usepackage{multirow}
\usepackage[normalem]{ulem}
\usepackage{lipsum}
\usepackage{ytableau}
\usepackage{youngtab}

\usepackage[mathscr]{eucal}
\usepackage{color}
\usepackage{hyperref}
\newtheorem{theorem}{Theorem}

\ytableausetup{boxsize=0.8em}

\newcommand{\Tr}{\operatorname{Tr}}

\newcommand{\bra}[1]{\langle #1|}
\newcommand{\ket}[1]{|#1\rangle}

\definecolor{cadmiumgreen}{rgb}{0.0, 0.42, 0.24}

\newcommand{\SU}{\operatorname{SU}}

\begin{document}

% Main document content 

\title{Quantum Advantage in Identifying the Parity of Permutations with Certainty}
\author{A. Diebra$^{1}$, S. Llorens$^{1}$, D. Gonz\'alez-Lociga$^1$, A. Rico$^1$, J. Calsamiglia$^1$, M. Hillery$^{2}$, and E. Bagan$^{1}$}

\affiliation{$^{1}$F\'{i}sica Te\`{o}rica: Informaci\'{o} i Fen\`{o}mens Qu\`antics, Universitat Aut\`{o}noma de Barcelona, 08193 Bellaterra (Barcelona), Spain
}
\affiliation{$^{2}$Department of Physics and Astronomy, Hunter College of CUNY, 695 Park Avenue, New York, NY 10065 USA and Physics Program, Graduate Center of CUNY, 365 Fifth Avenue, New York, NY 10016 USA
}

\begin{abstract}
We establish a sharp quantum advantage in determining the parity (even/odd) of an unknown permutation applied to any number $n \ge 3$ of particles. Classically, this is impossible with fewer than~$n$ labels, being that the success is limited to random guessing. Quantum mechanics does it with certainty with as few as $\lceil \sqrt{n}\, \rceil$ distinguishable states per particle, thanks to entanglement. Below this threshold, not even quantum mechanics helps: both classical and quantum success are limited to random guessing. For small $n$, we provide explicit expressions for states that ensure perfect parity identification. We also assess the minimum entanglement these states need to carry, finding it to be close to maximal, and even maximal in some cases. The task requires no oracles or contrived setups and provides a simple, rigorous example of genuine quantum advantage.
\end{abstract}

\maketitle

\noindent\textit{Introduction}---One of the defining goals of quantum information science is to identify tasks that quantum systems can perform better than classical ones, a phenomenon broadly known as quantum advantage. Over the past two decades, many such examples have emerged: exponential speedups relative to the best known classical algorithms, as in Shor’s factoring~\cite{Shor1997}; reductions in communication complexity~\cite{Buhrman2010}, e.g., random access codes~\cite{PhysRevLett.114.170502} and quantum fingerprinting~\cite{buhrman2001quantum}; improvements in learning efficiency~\cite{huang2022quantum}; or exponential query savings in oracle-based problems, such as Simon’s~\cite{Simon1997} and the Bernstein--Vazirani~\cite{doi:10.1137/S0097539796300921,naseri2022entanglement} algorithms—both special cases of the \emph{Abelian} hidden subgroup problem.
The recent experimental achievement of quantum supremacy~\cite{Arute2019,Preskill2012} can be viewed as an extreme case of quantum advantage, where a quantum device performs a computational task beyond the reach of classical supercomputers.

While a significant portion of the literature has focused on quantum algorithms and computational tasks, quantum advantage also arises in a variety of physically motivated settings, from quantum communication and metrology to nonlocal correlations. These instances highlight how quantum resources such as coherence~\cite{Streltsov2017}, contextuality~\cite{amaral2019resource,gupta2023quantum}, and entanglement~\cite{plenio2014introduction} can yield capabilities fundamentally unattainable by classical systems. For example, entanglement-enhanced metrology enables Heisenberg-limited precision~\cite{Giovannetti2011}, and quantum communication protocols such as teleportation~\cite{Bennett1993} and dense coding~\cite{Bennett1992} enable tasks that are fundamentally impossible using classical resources alone.

In this Letter, we propose a problem that is elementary to state, physically natural, and yet exhibits a sharp separation between classical and quantum capabilities. It requires no complex communication setup, no oracle for function evaluation, and no computational promises---just a set of $n$ particles, subject to a hidden permutation, and the simple binary question: \emph{is it even or~odd?}

Swapping is arguably the most primitive and universal transformation one can perform on a set of objects. It applies regardless of their nature; it is simply a rearrangement. Classically, if each particle carries a distinct label---e.g., $n$ different colors---one can perfectly reconstruct the permutation and read off its parity. But with even one fewer label, perfect parity identification becomes impossible: every permutation has an opposite-parity counterpart producing the same label arrangement, and the best one can do is guess, with success probability $P_{\rm s} = 1/2$.

Quantum systems, however, allow us to do fundamentally better. By preparing the $n$ particles in a suitable initial entangled state $|\psi_e\rangle$ of qudits (each of local dimension $d$), we show that

\begin{theorem}\label{theorem}
Perfect parity identification ($P_{\rm s} = 1$) is achievable if
\begin{equation}
d \geq d_{\rm min} := \left\lceil \sqrt{n}\, \right\rceil.
\label{e:theor}
\end{equation}
Below this threshold, assuming all permutations are equally likely, parity remains indistinguishable and %random guessing is the best one can do 
one cannot do better than random guessing ($P_{\rm s} = 1/2$).
\end{theorem}
Thus, from a quantum system in which each of the $n$ constituents carries quadratically fewer distinguishable states than classically required, one can nonetheless extract the parity of the hidden permutation with certainty, thanks to entanglement---for local dimension below $d_{\rm min}$, not even quantum resources help. If Eq.~(\ref{e:theor}) holds, a parity-detecting state~$|\psi_e\rangle$ exists, and the subspaces spanned by the action of the even and odd permutations on $|\psi_e\rangle$ are mutually orthogonal, allowing parity identification by a projective measurement.

Theorem~\ref{theorem} makes our protocol a clear-cut example of genuine quantum advantage: the task is naturally defined, with a rigorously provable classical limitation, and, as will be shown, its quantum solution relies on fundamental structural properties of multipartite entangled states under permutation symmetry.

We emphasize that our result, summarized in Theorem~\ref{theorem}, is \emph{nonasymptotic}: it holds exactly for every~$n$, not just in the large-$n$ limit. This contrasts with other permutation-based protocols, such as those introduced in~\cite{korff2004quantum} and extended in~\cite{PhysRevA.71.012326}, where certainty is achieved~only asymptotically. %In~\cite{korff2004quantum}, asymptotic zero-error identification of the applied permutation (not just its parity) is achievable iff $d \ge n/\mathrm{e}$, where $\mathrm{e}$ is Euler's constant, while~\cite{PhysRevA.71.012326} shows that the same asymptotic performance can be attained with $d \sim \sqrt{n}$ using ancillary systems.

Our protocol  also bears some resemblance with the algorithm introduced in~\cite{gedik2015computational} and experimentally demonstrated in~\cite{dogra2014determining,wang2015demonstration,dai2018demonstration}. Although it involves a single quantum register (a single qudit), with the permutation applied to its internal basis states via a quantum oracle, the state used for perfect parity identification in the qutrit case can be mapped to the three-qubit version of our general $n$-qudit state~$|\psi_e\rangle$~\cite{supp_mat,denker2025chiral}:
\begin{equation}
|\psi_e\rangle = |011\rangle + \zeta_3 |101\rangle + \zeta_3^2 |110\rangle.
\label{psie for n=3}
\end{equation}
(Throughout this Letter, states are given unnormalized)
Here $\zeta_N := {\rm e}^{2\pi i/N}$ is the $N$th root of unity, and we adopt the standard notation \mbox{$|i_1 i_2 \dots\rangle = |i_1\rangle \otimes |i_2\rangle \otimes \cdots$}, with $i_p \in \{0,1,\dots,d-1\}$ for $p = 1,2,\dots,n$. One can easily verify that $\sigma_{\rm odd}|\psi_e\rangle$ is orthogonal to \mbox{$\sigma_{\rm even}|\psi_e\rangle \propto |\psi_e\rangle$} for any odd, respectively even, permutation, 
where %the action of a permutation $\sigma$ is simply
\mbox{$\sigma |i_1 i_2 \cdots\rangle = |i_{\sigma^{-1}(1)} i_{\sigma^{-1}(2)} \cdots\rangle$}. 
%
%This can be verified directly in the three-qubit case, where $|\psi_e\rangle$ is given in Eq.~(\ref{psie for n=3}) below.
%
In the ququart case ($d=4$), however, the authors consider only a restricted subset of permutations, and the notion of parity in that context is not clearly defined. Since their scheme involves a single qudit, no entanglement is present, and the observed quantum advantage is attributed instead to contextuality.

%The reported quadratic advantage formally resembles that of Grover’s search algorithm, but the latter concerns time complexity (fewer iterations or oracle calls), whereas here it is related to space complexity, since our (deterministic) protocol is single shot and requires quadratically fewer distinct labels. Hence no substantive connection between the two can be stablished. This time–versus–space distinction also singles out our protocol from most quantum algorithms, including those for binary identification (like ours), e.g., Refs.~\cite{deutsch1992rapid,farhi1998limit,dorai2001implementation,gedik2015computational,dogra2014determining,wang2015demonstration,dai2018demonstration}.

The reported quadratic advantage is formally analogous to that of Grover’s search algorithm, but the analogy is limited. The latter achieves a quadratic reduction in time through fewer oracle queries, whereas our deterministic, single-shot protocol achieves a quadratic reduction in space, with fewer distinct labels. Hence, no substantive correspondence between the two protocols appears to exist. This time–versus–space distinction also singles out our protocol from most quantum algorithms, including those for binary identification (like ours), e.g., Refs.~\cite{deutsch1992rapid,farhi1998limit,dorai2001implementation,gedik2015computational,dogra2014determining,wang2015demonstration,dai2018demonstration}.

This Letter is organized as follows. We begin with three explicit examples to gradually introduce notation and highlight features that generalize to arbitrary~$n$.
We then prove Theorem~\ref{theorem}, using group and representation theory. %; details are deferred to the supplemental material~\cite{supp_mat}.
%Next, we show how to compute parity-detecting states~$|\psi_e\rangle$ for general $n$ and $d$, and give expressions for small $n$.
Finally, we assess the multipartite entanglement required by our protocol and close with a brief summary and outlook.

\noindent\textit{Example: Four qubits}---The case of four particles already reveals the ingredients needed for perfect parity identification. For concreteness, consider four spin-$\tfrac{1}{2}$ particles, on which some agent (Alice) acts by permuting their positions without affecting their spin states. One can verify~that  
\begin{multline}
|\psi_e\rangle = |0011\rangle + |1100\rangle \\
+ \zeta_3 \left(|0101\rangle + |1010\rangle\right)
+ \zeta_3^2 \left(|0110\rangle + |1001\rangle\right),
\label{M4}
\end{multline}
%written in the computational basis (
with $0$ and $1$ denoting spin-up and spin-down, respectively, transforms under permutations exactly as its three-qubit analog in Eq.~(\ref{psie for n=3}): even permutations preserve it up to a phase, while odd ones map it to an orthogonal state. Hence, parity identification for four qubits is accomplished by a projective measurement onto these (one-dimensional) orthogonal subspaces. We now show how Eq.~(\ref{M4}) can be derived.

The Hilbert space of the four spins, $(\mathbb{C}^2)^{\otimes 4}$, can be decomposed (via, e.g., the standard Clebsch–Gordan coefficients) into a direct sum of $\mathrm{SU}(2)$-invariant subspaces $\mathscr{H}_j$ labeled by the total spin $j$~\cite{hamermesh2012group}:
\begin{equation}
(\mathbb{C}^2)^{\otimes 4} \cong \left( \mathscr{K}_2 \otimes \mathscr{H}_2 \right) \oplus \left( \mathscr{K}_1 \otimes \mathscr{H}_1 \right) \oplus \left( \mathscr{K}_0 \otimes \mathscr{H}_0 \right),
\label{S-W 4}
\end{equation}
where the $\mathscr{K}_j$ factors account for the multiplicity of each spin sector: $\mathscr{K}_2 \cong \mathbb{C}$, $\mathscr{K}_1 \cong \mathbb{C}^3$, and $\mathscr{K}_0 \cong \mathbb{C}^2$ (corresponding to 1, 3, and 2 copies, respectively). The group $\mathrm{SU}(2)$ acts only on the spin factors $\mathscr{H}_j$ and trivially on the multiplicity spaces $\mathscr{K}_j$. 

In contrast, any reshuffling of the four qubits translates into a “rotation” within each $\mathscr{K}_j$. More precisely, these spaces carry irreducible representations (irreps, for short) of the symmetric group $S_4$, the group of all permutations~of four elements. We recall that ``irreducible" means that no proper subspace is left invariant by the group action, which we denote $\sigma|\psi\rangle$ for $\sigma\in S_4$, $|\psi\rangle\in(\mathbb{C}^2)^{\otimes 4}$. Since the agent's operations are limited to such permutations, she acts trivially on the spin sectors $\mathscr{H}_j$, which can thus be viewed as multiplicity spaces for the irreps of $S_4$. This is~the perspective we adopt throughout the Letter, reflecting the essence of Schur-Weyl duality.

If we restrict the agent’s actions to \emph{even permutations}, which themselves form the group $A_4$ ($\subset S_4$)~\cite{fulton2013representation}---the alternating group of degree 4---the representations carried by $\mathscr{K}_2$ and $\mathscr{K}_1$ remain irreducible.
The case of $\mathscr{K}_0$ is different. Under $A_4$, it decomposes into two orthogonal one-dimensional invariant subspaces, denoted $\mathscr{K}_{0a}$ and $\mathscr{K}_{0b}$, each carrying a distinct irrep of $A_4$: $\mathscr{K}_0 = \mathscr{K}_{0a} \oplus \mathscr{K}_{0b}$. Moreover, any \emph{odd} permutation $\sigma_{\rm odd}$ ($\in S_4\backslash A_4$) maps vectors in $\mathscr{K}_{0a}$ to $\mathscr{K}_{0b}$ and vice versa, which is why $\mathscr{K}_{0a}$ and $\mathscr{K}_{0b}$ are not invariant under the full group $S_4$ while $\mathscr{K}_{0a} \oplus \mathscr{K}_{0b}$ is. By preparing the initial state of the four-spin system as $|\psi_e\rangle \in \mathscr{K}_{0a} \otimes \mathscr{H}_0$, one guarantees that
\begin{equation}
 \sigma_{\rm odd} | \psi_e\rangle\quad  \perp\quad   \sigma_{\rm even} | \psi_e\rangle.
\label{psi sigma psi}
\end{equation}

The expression of $|\psi_e\rangle$ in Eq.~(\ref{M4}) can be readily obtained~\cite{supp_mat} as \mbox{$|\psi_e\rangle = \mathscr{P}_{0a} |0011\rangle$}, where $\mathscr{P}_{0a}$ denotes the projector onto the invariant subspace \mbox{$\mathscr{K}_{0a} \otimes \mathscr{H}_0$}~\cite{fulton2013representation} (we choose $|0011\rangle$ since it has magnetic quantum number \mbox{$m=0$} and thus a component in this subspace). This projector can be written in terms of the actions of the alternating group~$A_4$ on $(\mathbb{C}^2)^{\otimes 4}$ as
\mbox{$\mathscr{P}_{0a} = |A_4|^{-1} \sum_{\sigma \in A_4} \overline{\chi}_{0a}(\sigma)\, \sigma$}, 
where the bar stands for complex conjugation,  \mbox{$|A_4| = 12$} is the order of $A_4$ and $\chi_{0a}$ denotes the character of the irrep~$0a$. 
Projecting any other state with \mbox{$m=0$} not annihilated by $\mathscr{P}_{0a}$ would also yield a valid choice.
The Fourier-type combination of Dicke states in~Eq.~(\ref{M4})~\cite{higuchi2000entangled,kiesel2007}, known as the M4 (or D4) state, is a prominent example of multipartite entanglement~\cite{bastin2009}.

Before proceeding with the next example, we introduce a conventional labeling of the $\mathrm{SU}(d)$- and $S_n$-invariant subspaces 
$\mathscr{H}$ and $\mathscr{K}$ using partitions of $n$, or equivalently, Young diagrams (YDs) with $n$ boxes~\cite{fulton2013representation}. 
We denote by $\lambda = [\lambda_1, \lambda_2, \dots]$ ($\lambda_1 \ge \lambda_2 \ge \cdots \ge 0$, $\sum_p \lambda_p = n$) any such partition, 
where each $\lambda_p$ specifies the length of row~$p$ in the associated diagram, 
and use the notation $\lambda \vdash n$ to indicate that $\lambda$ is a valid partition of $n$, equivalently a YD with $n$ boxes. 
By $\ell(\lambda)$ we denote the number of rows of the associated YD, referred to as its “length.” 
Each $\lambda \vdash n$ corresponds to a distinct irrep of $S_n$, labeled as $\mathscr{K}_\lambda$ 
(from now on, we follow the standard practice of referring to “irrep” both for the invariant subspaces 
$\mathscr{K}$, $\mathscr{H}$ and the representations they carry). 

The irreps of $\mathrm{SU}(d)$ can also be labeled by YDs, with the condition $\ell(\lambda) \le d$~\cite{supp_mat}, 
though they may contain an arbitrary number of boxes—reflecting that there are infinitely many irreps of $\SU(d)$. 
For qubits, $\lambda = [\lambda_1, \lambda_2]$ and one has $j = (\lambda_1 - \lambda_2)/2$.

\begin{figure}[h]
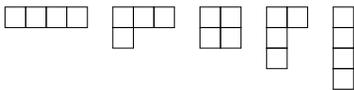

    \centering
    $$
    \ydiagram{4}\quad
    \ydiagram{3,1}\quad
    \ydiagram{2,2}\quad
    \ydiagram{2,1,1}\quad
    \ydiagram{1,1,1,1}
    $$
    \caption{The YDs corresponding to the five partitions of 4.}
    \label{fig:young-partitions-4}
\end{figure}
\noindent\textit{Example: Four qudits ($d>2$)}---This case reveals another key ingredient enabling zero-error parity identification. For $n=4$, the relevant YDs are shown in Fig.~\ref{fig:young-partitions-4}.
They correspond to the five partitions: $[4]$, $[3,1]$, $[2,2]=[2^2]$, $[2,1,1]=[2,1^2]$, and $[1,1,1,1]=[1^4]$.
For $d \ge 4$, all these YD-labeled irreps appear in the Schur–Weyl decomposition of~$(\mathbb{C}^d)^{\otimes n}$, analogous to that in Eq.~(\ref{S-W 4}). The subspaces $\mathscr{K}_{[4]}$ and $\mathscr{K}_{[1^4]}$ carry the fully symmetric (trivial: $\sigma \mapsto 1$ for all $\sigma \in S_n$) and fully antisymmetric (sign: $\sigma_{\rm even/odd} \mapsto \pm 1$) irreps of~$S_4$. We note that if $|\psi_{\rm sym}\rangle\in\mathscr{K}_{[4]}\otimes\mathscr{H}_{[4]}$ and $|\psi_{\rm ant}\rangle\in\mathscr{K}_{[1^4]}\otimes\mathscr{H}_{[1^4]}$ have the same norm, then, e.g.,
%
%\begin{equation}
$
|\psi_e\rangle = |\psi_{\rm sym}\rangle + |\psi_{\rm ant}\rangle
$
%\label{sym-ant}
%\end{equation}
%
also satisfies the orthogonality condition in Eq.~(\ref{psi sigma psi}), and enables perfect parity identification.

For $d=3$, however, the last YD is not allowed, as its length $\ell([1^4])=4$ exceeds $d=3$, and no fully antisymmetric state $|\psi_{\rm ant}\rangle$ can be constructed. Nonetheless, a state analogous to $|\psi_e\rangle$ in the previous paragraph % in Eq.~(\ref{sym-ant})
can still be assembled by combining the irreps $[3,1]$ and $[2,1^2]$ (the second and fourth YDs in Fig.~\ref{fig:young-partitions-4}).
These partitions, and their corresponding irreps, are known as conjugate to one another. Conjugate pairs of partitions, $\lambda$ and $\lambda^T$, can be easily recognized from their YD: they are obtained by transposing rows and columns, or equivalently, by reflecting the diagram along its main diagonal. The irreps they label satisfy the relation $D_{\lambda^T} \cong \mathrm{sign} \otimes D_{\lambda}$~\cite{fulton2013representation}, where $D_\lambda$ and $D_{\lambda^T}$ are the representation maps on $\mathscr{K}_{\lambda}$ and $\mathscr{K}_{\lambda^T}$ (and their matrices), respectively. In our example, this equivalence means that, by choosing bases appropriately, one actually has
$D_{[2,1^2]}(\sigma) = \mathrm{sign}(\sigma) D_{[3,1]}(\sigma)$ for all permutations. This holds in general:
\begin{equation}
D_{\lambda^T}(\sigma) = \mathrm{sign}(\sigma) D_{\lambda}(\sigma), \quad \mbox{for all $\sigma \in S_n$}.
\label{conjugate general}
\end{equation}
Hence, one can pick states (of equal norm) $|\psi_{\lambda}\rangle \in \mathscr{K}_\lambda \otimes \mathscr{H}_\lambda$ for $\lambda = [3,1], [2,1^2]$ so that
\begin{equation}
|\psi_e\rangle = |\psi_{[3,1]}\rangle + |\psi_{[2,1^2]}\rangle
\label{31+21^2}
\end{equation}
satisfies Eq.~(\ref{psi sigma psi}). %, thus enabling perfect parity identification. 
However, this case is slightly more involved than the qubit example because $\mathscr{K}_{[3,1]}$ and $\mathscr{K}_{[2,1^2]}$ are both 3-dimensional, and the measurement that identifies parity consists of projectors onto the spans of $\sigma_{\rm even}|\psi_e\rangle$ and $\sigma_{\rm odd}|\psi_e\rangle$, for all even and odd permutations.
The simple parity-detecting state
\begin{multline}
|\psi_e\rangle =  
|1000\rangle 
 +  |0100\rangle 
 +  |0010\rangle 
 -3  |0001\rangle 
 + 
\\
 \sqrt{2} \Big(
|0120\rangle 
 -  |0210\rangle 
      - |1020\rangle 
    + 
  \\[-.4em]
  |1200\rangle 
    + |2010\rangle 
    - |2100\rangle
\Big)
\label{conj_pairs_4}
\end{multline}
illustrates the construction~\cite{supp_mat}, but it is suboptimal, as it uses $3=d > d_{\rm min}=2$. Moreover, any (pure or mixed) state $\rho_e$ with $\mathrm{supp}(\rho_e) \subseteq \mathrm{span}\{\sigma_{\rm even}|\psi_e\rangle\}$ would likewise enable perfect parity identification.

The irrep $\mathscr{K}_{[2^2]}$ (third YD in Fig.~\ref{fig:young-partitions-4}) is called self-conjugate, since \mbox{$\lambda^T = \lambda$}. Under the action of $A_n$, the self-conjugate irreps split in two as $\mathscr{K}_{\lambda a}\oplus\mathscr{K}_{\lambda b}$ with $\mathscr{K}_{\lambda a}\stackrel{\sigma}{\longleftrightarrow} \mathscr{K}_{\lambda b}$ if $\sigma$ is odd, just as for qubits.  This can be used to enable parity identification exactly as in our first example, although there this was achieved with lower local dimensionality:~$d = 2$. Self-conjugate irreps and pairs of conjugate irreps are the only two mechanisms by which parity identification with certainty can be achieved, as we will prove after the next example.

\noindent\textit{Example: Five qutrits (\mbox{$d=3$})}---This is the first instance of a parity-detecting state with non-trivial coefficients (not just phases). The Schur-Weyl decomposition of~$(\mathbb{C}^3)^{\otimes 5}$ contains the self-conjugate irrep \mbox{$\mathscr{K}_{ [3,1^2]}$}, of dimension \mbox{$d_{[3,1^2]} = 6$}. Parity can be identified with certainty if the five qutrits are prepared in any pure state \mbox{$|\psi_e\rangle \in \mathscr{K}_{[3,1^2]a}\otimes \mathscr{H}_{[3,1^2]}$}, or in any mixed state $\rho_e$ with support entirely in this subspace.
Using the projector $\mathscr{P}_{[3,1^2]a}$ onto $\mathscr{K}_{[3,1^2]a}\otimes \mathscr{H}_{[3,1^2]}$ (analogous to $\mathscr{P}_{0a}$ in our first example), whose generic expression is~\cite{fulton2013representation}
\begin{equation}
\mathscr{P}_{\lambda a} = \frac{d_\lambda/2}{|A_n|} \sum_{\sigma \in A_n} \overline{\chi}_{\lambda a}(\sigma) \, \sigma ,
\label{proj}
\end{equation}
one can express any such state $|\psi_e\rangle$ in the computational basis as \mbox{$|\psi_e\rangle = |\psi_{[3,1^2] a}\rangle := \mathscr{P}_{[3,1^2] a} |\psi_0\rangle$}, where $|\psi_0\rangle$ is any state not annihilated by the projector (i.e., with non-vanishing components in the invariant subspace).
The~simple choice \mbox{$|\psi_0\rangle = |00012\rangle$} yields~\cite{supp_mat}:
\begin{multline}
|\psi_e\rangle=
3\big(
|00012\rangle - |00021\rangle
\big)
- \\
|00102\rangle + |00120\rangle + |00201\rangle - |00210\rangle 
-\\
 |01002\rangle + |01020\rangle - |10002\rangle + |10020\rangle 
+\\
 |02001\rangle - |02010\rangle + |20001\rangle - |20010\rangle
+\\
 \sqrt{5}\big(
|01200\rangle - |02100\rangle - |10200\rangle +\\
 |12000\rangle 
+ |20100\rangle - |21000\rangle
\big).
\label{self_5}
\end{multline}
These non-trivial coefficients reflect that $A_5$ is non-solvable~\cite{supp_mat}—the smallest alternating group with this property~\cite{conway1985atlas}.

\noindent\textit{Proof of Theorem~\ref{e:theor}}---Parity identification is an instance of binary discrimination between two quantum hypotheses:
\begin{equation}
\rho_0 = \frac{1}{|A_n|} \sum_{\sigma \in A_n} \sigma |\psi_e\rangle \langle \psi_e| \sigma^{-1}, \quad
\rho_1 = (12) \rho_0 (12),
\end{equation}
(note $\rho_1 = \sigma_{\rm odd} \rho_0 \sigma_{\rm odd}^{-1}$ for any \mbox{$\sigma_{\rm odd} \in S_n\! \setminus \!A_n$}). Perfect discrimination occurs iff $\Tr(\rho_0\, \rho_1) = 0$, i.e., iff the supports of the two density matrices are disjoint. If instead $\rho_0 = \rho_1$, then no information can be gained and random guessing is all one can do ($P_s = 1/2$).

The Hilbert space of $n$ qudits decomposes as~\cite{fulton2013representation}
\begin{equation}
(\mathbb{C}^d)^{\otimes n} \cong \bigoplus_{\substack{\lambda \vdash n \\ \ell(\lambda) \le d}} \mathscr{K}_\lambda \otimes \mathscr{H}_\lambda,
\label{decomp}
\end{equation}
which can be partitioned into three sectors, $R_{\rm i}$, $R_{\rm ii}$, and $R_{\rm iii}$, according to whether $\lambda$ satisfies:  
(i)~$\lambda = \lambda^T$;  
(ii)~\mbox{$\lambda \neq \lambda^T$} and $\ell(\lambda^T) \le d$;  
(iii)~$\ell(\lambda^T)>d$ [note that necessarily $\lambda\not=\lambda^T$ in this case, as $\ell(\lambda)\le d$].

%We will first assume, for simplicity, that for each $\lambda$ satisfying~(ii), the state $|\psi_e\rangle$ projects onto an equal number of copies of $\mathscr{K}_\lambda$ and $\mathscr{K}_{\lambda^T}$, so that effectively \mbox{$\dim\mathscr{H}_\lambda = \dim\mathscr{H}_{\lambda^T}$}.

Under the action of $A_n$ (restriction denoted $\downarrow_{A_n}$), in sector~(i) the irreps split: \mbox{$\mathscr{K}_\lambda \downarrow_{A_n} = \mathscr{K}_{\lambda a} \oplus \mathscr{K}_{\lambda b}$}.  
In sector~(ii), the irreps appear in conjugate pairs (see examples), and \mbox{$(\mathscr{K}_\lambda \oplus \mathscr{K}_{\lambda^T}) \downarrow_{A_n} = \mathscr{K}_\lambda \otimes \mathbb{C}^2$}, with $\ell(\lambda) < \ell(\lambda^T)$ [from Eq.~(\ref{conjugate general}), \mbox{$\mathscr{K}_\lambda=\mathscr{K}_\lambda\downarrow_{A_n}=\mathscr{K}_{\lambda^T}\downarrow_{A_n}$}, i.e., the same irrep appears twice].  
In sector~(iii), \mbox{$\mathscr{K}_\lambda \downarrow_{A_n} = \mathscr{K}_\lambda$}: the restriction has no effect. 

Accordingly, let us write $\rho_0 = \rho_0^{\rm (i)} + \rho_0^{\rm (ii)} + \rho_0^{\rm (iii)}$. Invoking Schur’s lemma, we find that these matrices are block-diagonal:
\begin{align}
\rho_0^{\rm (i)} &= \sum_{\lambda \in R_{\rm i}} \left( \frac{\openone_{\lambda a}}{d_\lambda / 2} \otimes \Phi^{\lambda a} + \frac{\openone_{\lambda b}}{d_\lambda / 2} \otimes \Phi^{\lambda b} \right), 
\label{rhoi} \\[-0.3em]
\rho_0^{\rm (ii)} &= \sum_{\lambda \in R_{\rm ii}} \frac{\openone_\lambda}{d_\lambda} \otimes \sum_{k=1}^{d_\lambda} |\phi^\lambda_k\rangle \langle \phi^\lambda_k|, 
\label{rhoii} \\
\rho_0^{\rm (iii)} &= \sum_{\lambda \in R_{\rm iii}} \frac{\openone_\lambda}{d_\lambda} \otimes \Phi^\lambda,
\label{rhoiii}
\end{align}
where $d_\lambda=\dim\mathscr{K}_\lambda$ and $\Phi^\lambda$, $\Phi^{\lambda a}$, $\Phi^{\lambda b}$ are operators on \mbox{$\mathscr{H}_\lambda=\mathscr{H}_{\lambda a}=\mathscr{H}_{\lambda b}$}, respectively. Similarly, $\openone_\lambda$, $\openone_{\lambda a}$, and~$\openone_{\lambda b}$ denote the identity operators on the corresponding $\mathscr{K}$-spaces, and
\begin{equation}
|\phi^\lambda_k\rangle = \sum_{p=\pm} |s_p\rangle |\phi^\lambda_{k,p}\rangle \in \mathbb{C}^2 \otimes \mathscr{H}_\lambda.
\end{equation}
Here $|s_{\pm}\rangle$ is an orthonormal basis of the multiplicity space $\mathbb{C}^2$ (upon which $A_n$ acts trivially) chosen so that $|s_{\pm}\rangle \stackrel{\raisebox{-.2ex}{\scriptsize$\sigma$}}{\longrightarrow} \pm |s_{\pm}\rangle$ for $\sigma \in S_n\! \setminus\! A_n$, and whose existence follows from Eq.~(\ref{conjugate general}).

These equations imply that

\noindent(a)~if either $\Phi^{\lambda a} = 0$ or $\Phi^{\lambda b} = 0$, then $\rho_0^{\rm (i)}$ and $\rho_1^{\rm (i)}$ have orthogonal supports;

\noindent(b)~after some algebra,
\begin{equation}
\Tr\!\left(\!\rho_0^{\rm(ii)} \!\rho_1^{\rm(ii)}\!\right)\! =\!\! \sum_{\lambda \in R_{\rm ii}} \sum_{k,l=1}^{d_\lambda}\! \frac{\left| \langle \phi^\lambda_{k,+} | \phi^\lambda_{l,+} \rangle \!-\! \langle \phi^\lambda_{k,-} | \phi^\lambda_{l,-} \rangle \right|^2\!\!}{d_\lambda}.
\end{equation}
If \mbox{$\langle \phi^\lambda_{k,+} | \phi^\lambda_{l,+} \rangle = \langle \phi^\lambda_{k,-} | \phi^\lambda_{l,-} \rangle$} for all $\lambda$, $k$, $l$, this trace vanishes and $\rho_0^{\mbox{{\tiny(}{\scriptsize ii}{\tiny)}}}$ and $\rho_1^{\mbox{{\tiny(}{\scriptsize ii}{\tiny)}}}$ have orthogonal supports [Eq.~(\ref{31+21^2}) follows from a particular case of this condition].

\noindent(c)~$\rho_0^{\mbox{{\tiny(}{\scriptsize iii}{\tiny)}}}$ is invariant under all of $S_n$, so $\rho_0^{\mbox{{\tiny(}{\scriptsize iii}{\tiny)}}} = \rho_1^{\mbox{{\tiny(}{\scriptsize iii}{\tiny)}}}$.

%In general, our simplifying assumption does not hold and there is a number \mbox{$|\dim\mathscr{H}_{\lambda}-\dim\mathscr{H}_{\lambda^T}|>0$} of copies of $\mathscr{K}_\lambda$ or $\mathscr{K}_{\lambda^T}$ that cannot be paired, just as in case~(iii). For this reason, the components of $|\psi_e\rangle$ on these subspaces give rise to an additional contribution to $\rho_0$ that satisfies~(c) and is included in $\rho_0^{\mbox{{\tiny(}{\scriptsize iii}{\tiny)}}}$.

By inspecting the shape of the YDs labeling the irreps
 in decomposition~(\ref{decomp}), it is straightforward to see that conditions~(i) and/or~(ii) hold if Eq.~(\ref{e:theor}) is satisfied. Then, from~(a) and~(b), there exist states $|\psi_e\rangle$ that enable  parity identification with certainty.
If $d < d_{\rm min}$, conditions~(i) and~(ii) fail and $\rho_0^{\mbox{{\tiny(}{\scriptsize i}{\tiny)}}} = \rho_0^{\mbox{{\tiny(}{\scriptsize ii}{\tiny)}}} = 0$. Thus, from~(c), $\rho_0 = \rho_1$, making random guessing the only option ($P_{\rm s} = 1/2$).~$\blacksquare$

\noindent\textit{How much entangled does a parity-detecting state need to be?}---To address this question, we use the geometric measure of entanglement (GME)~\cite{Wei2003geometric,Lisa2025quantifyin}. For a multipartite pure state $|\psi\rangle$, it is defined as
$
E(\ket{\psi}) = 1 - \max_{\ket{\phi} \in \mathrm{SEP}} |\langle \phi | \psi \rangle|^2
$,
where the maximization is over the set $\mathrm{SEP}$ of all separable states~\mbox{$\ket{\phi} = \bigotimes_{i=1}^n \ket{\phi_i}$}. Roughly speaking, $E(\ket{\psi})$ quantifies the “distance” from $\ket{\psi}$ to its closest separable state. The GME extends to mixed states via the standard convex roof construction.

Focusing on the optimal cases (\mbox{$d = d_{\rm min}$}) for \mbox{$n=3$}, $4$,~$5$, the state $\ket{\psi_e}$ can be any vector in $\mathscr{K}_{\lambda a} \otimes \mathscr{H}_{\lambda}$, where $\lambda = [2,1],\, [2,2],\, [3,1^2]$ respectively. The smallest amount of entanglement $|\psi_e\rangle$ requires can then be computed as~\cite{supp_mat}
\begin{equation}
    E_{\lambda a} = 1 - \max_{\ket{\phi} \in \mathrm{SEP}} \bra{\phi} \mathscr{P}_{\lambda a} \ket{\phi},
\end{equation}
where \mbox{$\mathscr{P}_{\lambda a}$} is the projector onto irrep \mbox{$\mathscr{K}_{\lambda a} \otimes \mathscr{H}_{\lambda}$}.
%
%since the spectral representation of the projector $\mathscr{P}_{\lambda a}$ is {\em any} uniform combination of orthogonal states spanning \mbox{$\mathscr{K}_{\lambda a} \otimes \mathscr{H}_{\lambda}$}.  
If \mbox{$\ket{\phi^*} \in \mathrm{SEP}$} maximizes \mbox{$\bra{\phi} \mathscr{P}_{\lambda a} \ket{\phi}$}, then a parity-detecting state with minimal entanglement is
\mbox{$ 
\ket{\psi_e} = \mathscr{P}_{\lambda a} \ket{\phi^*},
$}
which, once normalized, has a GME that lower bounds the entanglement of any parity-detecting state $\rho_e$ (pure or mixed) supported on this subspace~\cite{supp_mat}.

Table~\ref{tab:Esign} shows lower bounds for $E_{\lambda a}$, together with reference (greatest known) values of the geometric measure of entanglement, $E_{\rm max}$, for 3, 4, and 5 particles~\cite{Steinberg2024maxGME,denker2025chiral}. The bounds are computed numerically using symmetric extension~\cite{DPS2004,NavascuesDPS2009} while exploiting $\SU(d)$ symmetry. In all cases, $d=d_{\rm min}$; a dagger $({}^\dagger)$ indicates proven maxima. One sees that the entanglement required is maximal for $n=3,4$, and close to the known maximum for $n=5$.

\begin{table}[htb]
    \centering
    \begin{tabular}{c c c c c}
    \hline\hline
        $n$ & \quad $3$ & \quad $4$ & \quad $5$ \\
         \hline 
        $E_{\lambda a}$ & \quad $5/9\phantom{{}^\dagger}$ & \quad $7/9\phantom{{}^\dagger}$ & \quad $17/20$ \\
        $E_{\max}$ & \quad $5/9{}^\dagger$ & \quad $7/9{}^\dagger$ & \quad $\approx 0.96$ \\
        \hline\hline
    \end{tabular}
   \caption{Minimum entanglement required for parity identification with certainty (top row) versus known maximal entanglement for the same number of particles (bottom row).}
    \label{tab:Esign}
\end{table}

\noindent\textit{Summary and outlook}---We have presented a protocol that determines the parity of an unknown permutation on $n$ particles with certainty using as few as $\lceil \sqrt{n}\, \rceil$ distinguishable states per particle, far below the classical requirement of $n$. We provided explicit states for small~$n$ and quantified the minimum entanglement required, finding it close to maximal. This Letter reveals a sharp, provable quantum advantage that relies only on permuting particles, without invoking any specific dynamics. Note that our protocol requires no ancillas, nor can their inclusion reduce the minimum local dimension, as increasing the multiplicities has no effect here.

Future work should explore practical implementations, including preparation of the required states using experimentally accessible Hamiltonians (via ground-state engineering or dynamical protocols) or quantum circuits, along with design of the corresponding measurements. Assessing the robustness of the $\sqrt{n}$ quantum advantage under noise will also be essential. Finally, extensions to other symmetry groups may reveal further instances of clear quantum advantage~\cite{future_work}.

%\noindent\textit{Acknowledgments}. We thank R.~Mu\~noz-Tapia~for~discussions. 
%This work was supported by MCIN with funding from the EU's Next\-GenerationEU \mbox{(PRTR-C17.I1)}, the Generalitat de Catalunya, and the Spanish MINECO-TD's QUANTUM ENIA project ``Quantum Spain", funded by the EU's Recovery, Transformation and Resilience Plan – NextGenerationEU, under the “Digital Spain 2026 Agenda”. Additional support comes from grant PID2022-141283NB-I00, funded by \mbox{MICIU/AEI/10.13039/501100011033}. J.C. acknowledges support from the ICREA Academia programme. A.R. from MICIN (PID2022-139099NB-I00), with FEDER co-funding, and A.D. from the Spanish Ministry of Science and Innovation (FPU23/02763).
%
%-----------------

\vspace{-.1em}

\noindent\textit{Acknowledgments}---\mbox{This work was supported by MCIN} with funding from the EU's Next\-GenerationEU \mbox{(PRTR-C17.I1)}, the Generalitat de Catalunya, and the Spanish MINECO-TD through the QUANTUM ENIA project ``Quantum Spain", funded by the EU's Recovery, Transformation and Resilience Plan – NextGenerationEU under the “Digital Spain 2026 Agenda”. Additional support comes from grant PID2022-141283NB-I00 (MICIU/AEI/10.13039/501100011033). J.C. acknowledges support from ICREA Academia. A.R. was partially supported by MICIN (PID2022-139099NB-I00, FEDER co-funding), and A.D. by MICIN (FPU23/02763).

\textit{Data availability}---The data that supports the findings of this article are openly \cite{AlbertGit} in this Letter.

\bibliography{bibliography} 

\cleardoublepage 

\setcounter{section}{0} 
\setcounter{equation}{0} 
\setcounter{figure}{0} 
\setcounter{table}{0} 
\setcounter{page}{1}

\renewcommand{\theequation}{S\arabic{equation}}

\title{SUPPLEMENTAL MATERIAL: \\ Quantum Advantage in Identifying the Parity of Permutations with Certainty}
\author{A. Diebra$^{1}$, S. Llorens$^{1}$, D. Gonz\'alez-Lociga$^1$, A. Rico$^1$, J. Calsamiglia$^1$, M. Hillery$^{2}$, and E. Bagan$^{1}$}

\affiliation{$^{1}$F\'{i}sica Te\`{o}rica: Informaci\'{o} i Fen\`{o}mens Qu\`antics, Universitat Aut\`{o}noma de Barcelona, 08193 Bellaterra (Barcelona), Spain
}
\affiliation{$^{2}$Department of Physics and Astronomy, Hunter College of CUNY, 695 Park Avenue, New York, NY 10065 USA and Physics Program, Graduate Center of CUNY, 365 Fifth Avenue, New York, NY 10016 USA
}

\begin{abstract}
These supplemental notes provide the necessary group-theoretical background on the symmetric and alternating groups, including their conjugacy classes, irreducible representations, character tables, Schur–Weyl duality, and the construction of orthogonal projectors via generalized Young symmetrizers. We derive the parity-detecting states in the computational basis for the cases discussed in the main text and detail the numerical methods used to bound their entanglement via the geometric measure, using semidefinite programming relaxations.
\end{abstract}

%\pacs{03.67.-a, 03.65.Ta,42.50.-p }
\maketitle

\onecolumngrid

\section{\boldmath The groups $S_n$ and $A_n$ in a nutshell}

This brief overview of the symmetric group $S_n$, the alternating group $A_n$, and their representation theory (with occasional references to general groups $G$) is loosely inspired by~\cite{fulton2013representation,hamermesh2012group,wigner2012group,serre1977linear} and is included to keep these supplemental notes self-contained.

\subsection{\boldmath The symmetric group $S_n$}

The {\em symmetric group}~$S_n$ is the group of all permutations of~$n$ objects. Its elements can be represented as bijections~$\sigma: \{1,2,\dots,n\} \to \{1,2,\dots,n\}$, with group operation given by composition: applying~$\sigma$ followed by~$\tau$ yields the permutation~$\tau\sigma$. Its order (number of elements) is $|S_n| = n!$, and its identity element is denoted by $e$.

Any permutation $\sigma \in S_n$ can be expressed as a product of cycles. A {\em cycle} $(a_1, a_2, \dots, a_k)$ permutes elements cyclically: $a_1 \to a_2$, $a_2 \to a_3$, $\dots$, $a_k \to a_1$, leaving all other elements fixed. Any permutation can be written uniquely (up to order) as a product of {\em disjoint cycles}, i.e., cycles with no elements in common.

Any permutation can also be decomposed (non-uniquely) as a product of {\em transpositions}, i.e., 2-cycles $(a,b)$ that swap~$a$ and~$b$. The {\em parity} (evenness or oddness) of a permutation is defined by whether it can be expressed as an even or odd number of transpositions; this parity is well-defined (independent of the chosen decomposition).

\subsubsection{\boldmath Conjugacy classes}

The {\em conjugacy class} of an element~$g$ in a group~$G$ is the set of all elements $hgh^{-1}$ with $h \in G$; in~$S_n$, two permutations are conjugate if and only if they have the same cycle structure when written as a product of disjoint cycles.

Conjugacy classes in~$S_n$ are in one-to-one correspondence with partitions of~$n$, or equivalently, Young diagrams with~$n$ boxes. A {\em partition $\lambda = [\lambda_1, \lambda_2, \dots, \lambda_r]$ of~$n$} is a sequence of positive integers in non-increasing order summing to~$n$. The {\em Young diagram} (YD) of $\lambda$ is a left-justified arrangement of boxes with $\lambda_p$ boxes in row~$p$. Each YD provides a graphical depiction of a partition as rows of decreasing length. In this correspondence, the parts of the partition (or equivalently, the lengths of the rows of its YD) match the lengths of the disjoint cycles in the permutations of that conjugacy class---for example, $(1,2,4)(5,6)(3)\in S_6$ has cycle type $[3,2,1]$, corresponding to a 3-cycle, a 2-cycle, and a 1-cycle (with 1-cycles, such as $(3)$, typically omitted from the notation), whose YD is:
$$
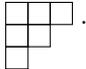
\ydiagram{3,2,1}\;.
$$
Thanks to this one-to-one correspondence, conjugacy classes in~$S_n$ can be conveniently labeled by partitions or their corresponding YDs.

We denote by $C_\lambda$ the conjugacy class corresponding to partition $\lambda$, though often we will simply refer to it as conjugacy class $\lambda$. Its size is given by
\begin{equation}
|C_\lambda| = \frac{n!}{\prod_p m_p! \lambda_p^{m_p}},
\end{equation}
with $m_p$ denoting the multiplicity of the part $\lambda_p$ in $\lambda$ (i.e., the number of rows of length $\lambda_p$ in the associated YD).

\subsubsection{\boldmath Representations}

A \emph{representation} of a group $G$ is a homomorphism from $G$ to the group of invertible linear operators on a vector space~$\mathscr{V}$. Since we are interested in physical systems, where $\mathscr{V}$ is a Hilbert space, we restrict attention to \emph{unitary} representations.
A representation is {\em irreducible} if $\mathscr{V}$ has no proper, nontrivial subspace invariant under all group elements. As is common practice, we will use the term ``representation" to refer to both the operators on $\mathscr{V}$ (or their matrices in a given basis) and the space $\mathscr{V}$ itself. The operators (or matrices) will be denoted $D(g)$ for $g \in G$. Accordingly, we will sometimes denote a representation by the pair $(\mathscr{V},D)$ for extra precision.
Any finite group $G$ has as many irreducible representations as conjugacy classes.

For $S_n$, the irreducible representations (irreps) are also in one-to-one correspondence with partitions of $n$ or YDs with $n$ boxes. We denote by $\mathscr{K}_\lambda$ the irrep labeled by partition/YD $\lambda$, though usually we will refer to $(\mathscr{K}_\lambda,D^\lambda)$ simply as ``irrep $\lambda$". Its dimension, $d_\lambda=\dim \mathscr{K}_\lambda=\Tr\big(D^\lambda(e)\big)$, is given by the {\em hook-length formula}:
\begin{equation}
d_\lambda = \frac{n!}{\prod_{b \in \lambda} h(b)},
\label{hook}
\end{equation}
where $h(b)$ is the hook length of box $b$ in the YD. The {\em hook length} of a box is the total number of boxes directly to its right and directly below it, plus one (counting the box itself). 

As for any finite group, any representation $(\mathscr{K}, D)$ of $S_n$ can be decomposed as a direct sum of irreps:
\begin{equation}
\mathscr{K} = \bigoplus_{\lambda} \mathscr{K}_\lambda \otimes \mathbb{C}^{m_\lambda},
\end{equation}
where $\mathbb{C}^{m_\lambda}$ accounts for the multiplicity, that is, the number of copies of $\mathscr{K}_\lambda$ in the decomposition.
%
%For $S_n$, the irreducible representations (irreps) are also in one-to-one correspondence with partitions of $n$ or YDs with $n$ boxes. We denote by $\mathscr{K}_\lambda$ the irrep labeled by partition/YD $\lambda$. Its dimension is given by the hook-length formula:
%\begin{equation}
%d_\lambda = \frac{n!}{\prod_{b \in \lambda} h(b)},
%\end{equation}
%where $h(b)$ is the hook length of box $b$ in the YD. The hook length of a box is the total number of boxes directly to its right and directly below it, plus one (counting the box itself). 
%
%As for any finite group, any representation $(\mathscr{K}, D)$ of $S_n$ can be decomposed as a direct sum of irreps:
%%
%\begin{equation}
%\mathscr{K} = \bigoplus_{\lambda} \mathscr{K}_\lambda \otimes \mathbb{C}^{m_\lambda},
%\end{equation}
%%
%with $m_\lambda$ the multiplicity of irrep $\lambda$ in the decomposition. 
This means that there is a basis of $\mathscr{K}$ such that
\begin{equation}
D(\sigma) = \sum_\lambda D^\lambda(\sigma) \otimes \openone_{m_\lambda},\qquad  \sigma\in S_n,
\end{equation}
and $\openone_{m_\lambda}$ denotes the identity operator on $\mathbb{C}^{m_\lambda}$. Note that $S_n$ acts trivially on the multiplicity spaces.

\subsubsection{\boldmath Characters and character table}

Given a representation $(\mathscr{V}, D)$ of a finite group $G$, its {\em character} is the function $\chi: G \to \mathbb{C}$ defined by
\begin{equation}
\chi(g) = \Tr D(g).
\end{equation}
Note that characters are constant on conjugacy classes. Characters provide a powerful tool for studying representations; for instance, the character of a direct sum is the sum of the characters, and two irreps are equivalent iff their characters are the same.

The {\em character table} of a group $G$ is a square array whose rows correspond to {\em irreducible characters} (i.e., the characters of the irreps), denoted $\chi^\alpha$, and columns to conjugacy classes. For a generic group, it has the form shown in Table~\ref{CT gen}.
\begin{table}[h]
$$
\renewcommand{\arraystretch}{1.25}
\begin{array}[t]{c|ccccc}
\text{Class} & C_1 & \cdots & C_\beta & \cdots & C_r \\
\text{Size}  & 1   & \cdots & |C_\beta| & \cdots & |C_r| \\
\hline
\chi^{1}       & 1         & \cdots & 1 & \cdots & 1 \\
\vdots         & \vdots    &        & \vdots & & \vdots \\
\chi^{\alpha}  & d_\alpha  & \cdots & \chi^{\alpha}(C_\beta) & \cdots & \chi^{\alpha}(C_r) \\
\vdots         & \vdots    &        & \vdots & & \vdots \\
\chi^{r}       & d_r       & \cdots & \chi^{r}(C_\beta) & \cdots & \chi^{r}(C_r) \\
\end{array}
\vspace{.5em}
\renewcommand{\arraystretch}{1.25}
$$
\caption{Character table of a generic finite group $G$.\label{CT gen}}
\end{table}
There, $r$ is the number of conjugacy classes in $G$, and $C_1 = \{e\}$ is the class of the identity. The irrep labeled~$1$ is the {\em trivial representation}, $D^1(g) = 1$ for all $g \in G$, which exists for any group. $\chi^\alpha(C_\beta)$ denotes the character value on any element of $C_\beta$. The labels $\alpha$ and $\beta$ range over the same set of indices since $G$ has as many irreps as conjugacy classes. Note that $\chi^\alpha(C_1) = \chi^\alpha(e) = d_\alpha$, the dimension of irrep $\alpha$.

Irreducible characters, $\chi^\alpha$, form an orthonormal basis for the space of class functions on~$G$, with inner product
\begin{equation}
\langle \chi^\alpha,\chi^\beta\rangle := \frac{1}{|G|} \sum_{g \in G}\, \overline{ \chi^\alpha(g)}\; \chi^\beta(g) = \delta_{\alpha\beta},\quad  1\le \alpha,\beta\le r ,
\label{ort char}
\end{equation}
where the bar ($\overline{\phantom{m}}$) denotes complex conjugation.
Equivalently, summing over all conjugacy classes~$C_\gamma$,
\begin{equation}
\langle \chi^\alpha,\chi^\beta\rangle = \sum_{\gamma=1}^r \frac{|C_\gamma|}{|G|}\; \overline{\chi^\alpha(C_\gamma)} \;\chi^\beta(C_\gamma) = \delta_{\alpha\beta},
\quad  1\le \alpha,\beta\le r.
\label{orth rel 1}
\end{equation}
These relations can be used, e.g., to compute multiplicities in a decomposition.

There is also a dual orthogonality relation for the columns of the character table:
\begin{equation}
\sum_{\alpha=1}^r \overline{\chi^\alpha(C_\gamma)}\; \chi^\alpha(C_\delta) = \frac{|G|}{|C_\gamma|} \delta_{\gamma\delta}, \quad  1\le\gamma,\delta\le r.
\label{orth rel 2}
\end{equation}
Both sets of relations can help complete the character table of a group, but they do {not} fully determine it, particularly for groups of large order.

The character tables of $S_4$ and $S_5$ are shown in Table~\ref{ct s4 & s5},
\begin{table}[b]
$$
\begin{array}{cc}
\renewcommand{\arraystretch}{1.25}
\begin{array}[t]{c|ccccc}
\text{Class} & [1^4] & [2,1^2] & [2^2] & [3,1] & [4] \\
\text{Size}  & 1     & 6       & 3     & 8     & 6 \\
\hline
\chi^{[4]}     & 1 &  1 &  1 &  1 &  1 \\
\chi^{[1^4]}   & 1 & -1 &  1 &  1 & -1 \\
\chi^{[3,1]}   & 3 &  1 & -1 &  0 & -1 \\
\chi^{[2,1^2]} & 3 & -1 & -1 &  0 &  1 \\
\chi^{[2^2]}   & 2 &  0 &  2 & -1 &  0 \\
\end{array}
\hspace{4em} 
&
\hspace{4em} 
\renewcommand{\arraystretch}{1.25}
\begin{array}[t]{c|ccccccc}
\text{Class} & [1^5] & [2,1^3] & [2^2,1] & [3,1^2] & [4,1] & [5] & [3,2] \\
\text{Size}  &   1   &   10    &   15    &   20   &   30    &   24   &  20 \\
\hline
\chi^{[5]}     &  1 &  1 &  1 &  1 &  1 &  1 &  1 \\
\chi^{[1^5]}   &  1 & -1 & 1 & 1 & -1 & 1 & -1 \\
\chi^{[4,1]}   &  4 & -2 & 0 & 1 &  0 & -1 & 1 \\
\chi^{[2,1^3]} &  4 &  2 & 0 & 1 &  0 & -1 & -1 \\
\chi^{[3,2]}   &  5 &  1 & 1 & -1 & -1 & 0 & 1 \\
\chi^{[2^2,1]} &  5 & -1 & 1 & -1 & 1 & 0 & -1 \\
\chi^{[3,1^2]} &  6 &  0 & -2 & 0 & 0 & 1 & 0 \\
\end{array}
\renewcommand{\arraystretch}{1}
\\
& \\[-.5em]
\fbox{$S_4$}   \hspace{4em} 
&
\hspace{4em} 
 \fbox{$S_5$}
\end{array}
$$
\caption{Character tables of $S_4$ and $S_5$.\label{ct s4 & s5}}
\end{table}
where  both irreducible characters and conjugacy classes are labeled with partitions/YDs. This table encodes all the information about the representation theory of~$S_n$; for instance, it allows one to compute decomposition multiplicities and inner products via orthogonality relations.
One can readily check that the character tables of $S_4$ and $S_5$ above fulfill these relations, Eqs.~(\ref{orth rel 1}) and~(\ref{orth rel 2}).

The first row in the character tables of $S_4$ and $S_5$ corresponds to the trivial representation.
The irrep in the second row is known as the {\em sign representation}, which exists for any $n\in\mathbb{N}$: $\operatorname{sign}(\sigma_{\rm even})=1$;  $\operatorname{sign}(\sigma_{\rm odd})=-1$ for any even, respectively odd, permutation. So, it gives the parity of $\sigma\in S_n$ for any $n\in\mathbb{N}$. 

Note that all rows of these character tables (i.e., all irreps) except the bottom one appear in pairs (1 and 2, 3 and~4,~\dots). Their associated YDs are related by transposition (a reflection along the main diagonal): $\lambda \rightarrow \lambda^T$, and we say that such irreps are {\em conjugate} to each other. The bottom row corresponds to a {\em self-conjugate} irrep, where~$\lambda = \lambda^T$.

In general, conjugate irreps satisfy
\begin{equation}
D^{\lambda^T} \cong \mathrm{sign}\otimes D^{\lambda},
\label{conjugation}
\end{equation}
meaning that if $\lambda \neq \lambda^T$, there exists a basis such that
\begin{equation}
D^{\lambda^T}(\sigma)= \mathrm{sign}(\sigma)\, D^{\lambda}(\sigma), \qquad \forall \,\sigma \in S_n.
\end{equation}
If $\lambda=\lambda^T$, such a basis cannot exist, as it would imply $D^\lambda(\sigma_{\rm odd})=0$, contradicting invertibility. Instead, Eq.~(\ref{conjugation}) yields an internal automorphism:
\begin{equation}
V\, D^{\lambda}(\sigma)\, V^\dagger \;=\; \mathrm{sign}(\sigma)\, D^{\lambda}(\sigma), \qquad \sigma\in S_n.
\label{U}
\end{equation}
These properties are reflected in the character tables of $S_4$ and $S_5$, since
\begin{equation}
\chi^{\lambda^T}(C_{\lambda'}) \;=\; \mathrm{sign}(C_{\lambda'})\, \chi^\lambda(C_{\lambda'}),
\end{equation}
for all irreps $\lambda$ and conjugacy classes $C_{\lambda'}$. In particular, if $\lambda=\lambda^T$, then $\chi^{\lambda}(C_{\lambda'})=0$ for all odd conjugacy classes.

\subsubsection{\boldmath Schur's lemma}

A fundamental result in the classification of irreducible representations is {\em Schur's lemma}. In its matrix form, it states that given two irreps, \( D^\alpha(g) \) and \( D^\beta(g) \), if a matrix \( T \) satisfies \( T D^\alpha(g) = D^\beta(g) T \) for all \( g \in \mathcal{G} \), then \( T \) is either invertible —implying that the two representations are {\em isomorphic}— or zero. Moreover, in the special case where both irreps coincide, \( \alpha= \beta \), i.e., when \( T \) commutes with \( D^\alpha(g) \) for all \( g \in \mathcal{G} \), \( T \) must be a scalar multiple of the identity in the corresponding irreducible subspace. In other words, \( T = \xi\, \openone_{d_\alpha} \) for some \( \xi \in \mathbb{C} \).

A direct consequence of Schur's lemma is that the matrix elements of two irreps satisfy the following orthogonality relation, known as the {\em Great Orthogonality Theorem},
\begin{equation}\label{eq_app:great_ortho_relation}
\sum_{g\in G}\left[D^\alpha(g)\right]_{k,l}\overline{\left[D^\beta(g)\right]}_{k',l'} = \delta_{\alpha,\beta}\delta_{k,k'}\delta_{l,l'}\frac{|G|}{d_\alpha}. 
\end{equation}
The character orthogonality relation in Eq.~(\ref{ort char}) follows directly from this theorem.

\subsection{\boldmath The action of $S_n$ and $\SU(d$); Schur–Weyl duality}

\subsubsection{Irreps of $\SU(d)$}

The irreps of $\SU(d)$ can also be labeled by YDs with at most $d$ rows but an arbitrary number of boxes, reflecting that $\SU(d)$ has infinitely many irreps. Their dimensions can be computed as
\begin{equation}
m_\lambda=\operatorname{dim}\mathscr{H}_\lambda=\frac{\prod_{p \geq q}\left(l_p-l_q\right)}{2!3!\cdots(d-1)!}, 
\qquad \text{where}\quad l_p := d-p+1+\lambda_p .
\label{m_lambda}
\end{equation}
Equivalently, $m_\lambda$ is the number of \emph{semi-standard Young tableaux} (SSYTs) of shape $\lambda$ with entries in $\{0,\dots,d-1\}$, i.e., fillings of the boxes of $\lambda$ with these numbers that are weakly increasing along rows and strictly increasing down columns. Each SSYT of shape $\lambda$ labels a basis element of $\mathscr{H}_\lambda$.
A convenient mnemonic for evaluating Eq.~(\ref{m_lambda}) is shown in Table~\ref{mnemonic}.

\begin{table}[htb]
$$
\begin{array}{rllllll}
&&&&&&\kern.8em\underline{\kern2em l_p \kern2em }\\[1em]
\raisebox{.2em}{\scriptsize$d+   $}\kern-.4em& 
                   \ydiagram{3} 
                                  &\kern-.35em\raisebox{.1em}{$\cdots$}&
                                                 \kern-.64em \ydiagram{2}&
                                                             \kern-.35em\raisebox{.1em}{$\cdots$}&
                                                                            \kern-.64em \ydiagram{2}&
                                                                                                   \raisebox{.2em}{\scriptsize$\kern-.2em = d+\lambda_1$} \\[-.34em]
\raisebox{.2em}{\scriptsize$d\!-\!1+$}\kern-.4em& 
                  \ydiagram{3}&
                                   \kern-.35em\raisebox{.1em}{$\cdots$} &
                                                   \kern-.64em \ydiagram{2} & 
                                                                       &
                                                                                    &
                                                                                                 \raisebox{.2em}{\scriptsize$\kern-.2em=d-1+\lambda_2$} \\[-.36em]
\vdots\kern.3em&\kern1.2em\vdots&&&&&    \\[.2em]                                                                                             
\raisebox{.2em}{\scriptsize$d\!-\!p\!+\!1+$}\kern-.4em& 
                  \ydiagram{3}&
                                 \kern-.35em\raisebox{.1em}{$\cdots$}  &
                                          &
                                                    &
                                                                 &
                                                                          \raisebox{.2em}{\scriptsize$\kern-.2em=d-p+1+\lambda_p$} \\[-.36em]
\vdots\kern.3em&\kern1.2em\vdots&&&&&    \\[.2em] 
\raisebox{.2em}{\scriptsize$2+$}\kern-.4em&\ydiagram{2}  &\kern-1em\raisebox{.1em}{$\cdots$}&&&& \raisebox{.2em}{\scriptsize$\kern-.2em=2+\lambda_{d-1}$} \\[-.34em]
\raisebox{.2em}{\scriptsize$1+$}\kern-.4em&\ydiagram{1} \kern.5em\raisebox{.1em}{$\cdots$} &&&&& \raisebox{.2em}{\scriptsize$\kern-.2em=1+\lambda_{d}$} \\[-.36em]
\end{array}
$$
\caption{Mnemonic for computing $l_p$ in Eq.~(\ref{m_lambda}). \label{mnemonic}}
\end{table}

Any diagram containing a column of length $d$ corresponds to a $\SU(d)$ representation equivalent to one with that column removed [as reflected in Eq.~(\ref{m_lambda})], since it carries a determinant factor of the $\SU(d)$ transformation, which equals unity. It is standard practice to remove such columns so that irreps are uniquely labeled by diagrams with at most $d{-}1$ rows. Here, however, we do \emph{not} follow this convention, in order to keep the labeling of $S_n$ and $\SU(d)$ irreps compatible in the next sections.

\subsubsection{Schur-Weyl duality}

Consider a system of $n$ qudits, each with Hilbert space $\mathbb{C}^d$. The total Hilbert space is the tensor product $(\mathbb{C}^d)^{\otimes n}$. The ``canonical'' basis of this space is the standard computational basis:
\begin{equation}
\ket{i_1 i_2 \cdots i_n} \equiv \ket{i_1} \otimes \ket{i_2} \otimes \cdots \otimes \ket{i_n}, \qquad i_l = 0, 1, \dots, d-1, \quad l = 1, 2, \dots, n.
\end{equation}
Both $\SU(d)$ and $S_n$ act naturally on $(\mathbb{C}^d)^{\otimes n}$, and their action is conveniently described in the computational basis. The group $\SU(d)$ acts diagonally:
\begin{equation}
U^{\otimes n} \ket{i_1 i_2 \cdots i_n} = \bigotimes_{l=1}^n U \ket{i_l}, \qquad U \in \SU(d),
\end{equation}
while $S_n$ acts by permuting tensor factors (inducing the \emph{tensor permutation representation}):
\begin{equation}
D^{\rm perm}(\sigma) \ket{i_1 i_2 \cdots i_n} 
= \ket{i_{\sigma^{-1}(1)} i_{\sigma^{-1}(2)} \cdots i_{\sigma^{-1}(n)}}, 
\qquad \sigma \in S_n.
\end{equation}
When no ambiguity arises, we will write $\sigma|\psi\rangle$ in place of $D^{\rm perm}(\sigma)|\psi\rangle$ for any $|\psi\rangle \in (\mathbb{C}^d)^{\otimes n}$.

The actions of $S_n$ and $\SU(d)$ commute and together generate a representation of $\SU(d) \times S_n$ on $(\mathbb{C}^d)^{\otimes n}$. %Since the irreps of $\SU(d)$ can also be labeled by YDs with at most $d$ rows (but arbitrary number of boxes, reflecting that $\SU(d)$ has infinitely many irreps), 
The decomposition of $(\mathbb{C}^d)^{\otimes n}$ into invariant subspaces can be expressed compactly (Schur–Weyl duality) as
\begin{equation}
(\mathbb{C}^d)^{\otimes n} \cong \bigoplus_{\substack{\lambda \vdash n \\ \ell(\lambda) \le d}} \mathscr{K}_\lambda \otimes \mathscr{H}_\lambda,
\label{SW-supp}
\end{equation}
where the sum runs over all partitions $\lambda$ of $n$, denoted $\lambda \vdash n$, and $\ell(\lambda)$ is the number of rows of $\lambda$, called its {\em length}. Here, $\mathscr{K}_\lambda$ carries, as above, the irrep of $S_n$ labeled by $\lambda$, and $\mathscr{H}_\lambda$ carries the irrep of $\SU(d)$ labeled by the same $\lambda$. The two factors are multiplicity spaces for each other: $S_n$ acts nontrivially only on $\mathscr{K}_\lambda$, and $\SU(d)$ acts nontrivially only on~$\mathscr{H}_\lambda$.

%The subgroup~$A_n$ (the alternating group on~$n$ elements) consists of all even permutations, i.e., those with positive parity. It is a normal subgroup of~$S_n$, with index~2 and order$|A_n|=n!/2$.
%
%In the representation theory context, irreducible representations (irreps) of~$S_n$ are labeled by partitions~$\lambda$ of~$n$, conveniently encoded as Young diagrams (YDs). Each YD provides a graphical depiction of a partition, with rows of decreasing length. The length of a partition,~$\ell(\lambda)$, is the number of rows in its diagram.

\subsection{Group algebra and projectors}

The {\em group algebra} $\mathbb{C}[G]$ of a finite group $G$ is the vector space spanned by formal linear combinations of its elements with complex coefficients, endowed with multiplication inherited from $G$ and extended by linearity. It has dimension~$|G|$, with basis $\{\,g \mid g \in G\,\}$.

Any representation $(\mathscr{V},D)$ of $G$ extends linearly to $\mathbb{C}[G]$:
\begin{equation}
D\Bigg(\sum_{g\in G} a_g g\Bigg) \;=\; \sum_{g\in G} a_g D(g) \;\in\; \mathrm{End}(\mathscr{V}), \qquad a_g \in \mathbb{C},
\end{equation}
where $\mathrm{End}(\mathscr{V})$ is the space of all linear operators (not necessarily invertible) on $\mathscr{V}$. 
In particular, every irrep of $G$ is also an irrep of $\mathbb{C}[G]$, and conversely. 

This correspondence allows one to construct {\em projectors onto the irrep subspaces} using characters:
\begin{equation}
\mathscr{P}_\alpha \;=\; \frac{d_\alpha}{|G|} 
\sum_{g \in G} \overline{\chi^\alpha(g)} \, D(g) 
\;\in\; \mathrm{End}(\mathscr{V}),
\label{proj gen}
\end{equation}
These projectors satisfy 
$\mathscr{P}_\alpha \mathscr{P}_\beta = \delta_{\alpha\beta}\,\mathscr{P}_\alpha$
and resolve the identity:
\begin{equation}
\sum_\alpha \mathscr{P}_\alpha \;=\; \openone 
\qquad \text{in } \mathrm{End}(\mathscr{V}).
\end{equation}

\noindent\textit{Proof.} Using the great orthogonality theorem, Eq.~(\ref{eq_app:great_ortho_relation}), one has
\begin{multline}
\sum_{g\in G}\overline{\chi^\alpha(g)}\,\chi^\beta(h^{-1}g)
=\sum_{k,l,s}\left(\sum_{g\in G}\overline{\left[D^\alpha(g)\right]_{k,k}}\left[D^\beta(g)\right]_{s,l}\right)\left[D^\beta(h^{-1})\right]_{l,s}
\\[.5em]
=\frac{|G|}{d_\alpha}\,\delta_{\alpha,\beta}\sum_{k,l,s}\delta_{k,s}\delta_{k,l}\left[D^\alpha(h^{-1})\right]_{l,s}
=\frac{|G|}{d_\alpha}\,\delta_{\alpha,\beta}\sum_{k}\left[D^\alpha(h^{-1})\right]_{k,k}
=\frac{|G|}{d_\alpha}\,\delta_{\alpha,\beta}\,\overline{\chi^{\alpha}(h)}.
\end{multline}
Then,
\begin{multline}
\mathscr{P}_\alpha\mathscr{P}_\beta
=\frac{d_\alpha d_\beta}{|G|^2}\sum_{g,g'\in G}\overline{\chi^\alpha(g)\chi^\beta(g')} \, D(g)D(g')
=\frac{d_\alpha d_\beta}{|G|^2}\sum_{g,g'\in G}\overline{\chi^\alpha(g)\chi^\beta(g')} \, D(gg')
\\[.5em]
=\frac{d_\alpha d_\beta}{|G|^2}\sum_{g,h\in G}\overline{\chi^\alpha(g)\chi^\beta(g^{-1}h)} \, D(h)
=\frac{d_\alpha d_\beta}{|G|^2}\sum_{h\in G}\left(\sum_{g\in G}\overline{\chi^\alpha(g)}\,\chi^\beta(h^{-1}g)\right) D(h)
\\[.5em]
=\delta_{\alpha,\beta}\,\frac{d_\alpha}{|G|}\sum_{h\in G}\overline{\chi^{\alpha}(h)} \, D(h)
=\delta_{\alpha,\beta}\,\mathscr{P}_\alpha,
\end{multline}
where we changed variables by setting $h := gg'$ in the second line.
This proves the orthogonality of the projectors.  

Next, we show completeness:
\begin{multline}
\sum_\alpha \mathscr{P}_\alpha
=\sum_\alpha \frac{\chi^\alpha(e)}{|G|}\sum_{g\in G}\overline{\chi^\alpha(g)}\,D(g)
=\sum_\alpha \frac{\chi^\alpha(e)}{|G|}\sum_{\beta}\sum_{g\in C_\beta}\overline{\chi^\alpha(g)}\,D(g)
\\[.5em]
=\sum_\alpha \frac{\chi^\alpha(C_1)}{|G|}\sum_{\beta}|C_\beta|\,\overline{\chi^\alpha(C_\beta)}\,D(C_\beta)
=\sum_{\beta} \frac{|C_\beta|}{|G|} \left( \sum_\alpha \overline{\chi^\alpha(C_\beta)}\,\chi^\alpha(C_1)\right) D(C_\beta)
\\[.5em]
= D(C_1)=D(e)=\openone \;\in \mathrm{End}(\mathscr{V}),
\end{multline}
where in the second line we used the orthogonality of the columns of the character table, Eq.~(\ref{orth rel 2}).  
This completes the proof~$\blacksquare$

When \mbox{$G=S_n$}, these projectors act naturally on $(\mathbb{C}^d)^{\otimes n}$ via the permutation representation, isolating the Schur--Weyl subspaces $\mathscr{K}_\lambda \otimes \mathscr{H}_\lambda$, and Eq.~\eqref{proj gen} becomes
\begin{equation}
\mathscr{P}_\lambda \;=\; \frac{d_\lambda}{n!} 
\sum_{\sigma \in S_n} \overline{\chi^\lambda(\sigma)} \, \sigma 
\;\in\; \mathrm{End}\!\left((\mathbb{C}^d)^{\otimes n}\right).
\label{proj Sn}
\end{equation}

\subsection{Standard and generalized Young symmetrizers}

For the symmetric group $S_n$, when explicit basis vectors of an irrep $\lambda$ are required, one can use {standard Young symmetrizers}. To introduce these, we first recall the notion of a {standard Young tableau}.

A {\em standard Young tableau} of shape $\lambda \vdash n$ is a filling of the YD corresponding to $\lambda$ with the integers $1,2,\dots,n$ such that entries strictly increase from left to right along each row and from top to bottom along each column.
The set of {standard Young tableaux} of shape $\lambda$ is denoted $\mathrm{SYT}(\lambda)$. Its cardinality $\lvert \mathrm{SYT}(\lambda)\rvert$ equals the dimension $d_\lambda$ of the irrep~$\lambda$, consistent with the hook-length formula in Eq.~(\ref{hook}).

\textit{Example.}
For $\lambda=[2,1^2]$, with YD $\ydiagram{2,1,1}$\;:
\begin{equation}
\mathrm{SYT}([2,1^2]) \;=\;
\left\{\, 
\raisebox{-1.1em}{\footnotesize\young(12,3,4)}, \;
\raisebox{-1.1em}{\footnotesize\young(13,2,4)}, \;
\raisebox{-1.1em}{\footnotesize\young(14,2,3)} 
\,\right\},
\end{equation}
so that $d_{[2,1^2]}=\lvert \mathrm{SYT}([2,1^2])\rvert=3$, matching Eq.~(\ref{hook}).

Given $T \in \mathrm{SYT}(\lambda)$, we define the \emph{row subgroup} and \emph{column subgroup} as
\begin{align}
R_T &:= \{ \sigma \in S_n \,\mid\, \sigma \ \text{permutes entries within each row of } T \}, \label{eq:RT}\\
C_T &:= \{ \sigma \in S_n \,\mid\, \sigma \ \text{permutes entries within each column of } T \}. \label{eq:CT}
\end{align}

\textit{Example.} For $T=\raisebox{-.6em}{\footnotesize\young(12,34)}$:
\begin{align}
R_T &= \{e,(12),(34),(12)(34)\}, \label{eq:RTex}\\
C_T &= \{e,(13),(24),(13)(24)\}. \label{eq:CTex}
\end{align}

From these, we define elements in the group algebra $\mathbb{C}[S_n]$:
\begin{align}
r_T &= \sum_{\sigma \in R_T} \sigma, \label{eq:rT}\\
c_T &= \sum_{\sigma \in C_T} \mathrm{sign}(\sigma)\, \sigma. \label{eq:cT}
\end{align}

The (normalized) \emph{standard Young symmetrizer} is then
\begin{equation}
Y_T \;:=\; \frac{d_\lambda}{n!}\, r_T \, c_T. \label{eq:YT}
\end{equation}
For fixed $\lambda$, the $Y_T$ associated with $T \in \mathrm{SYT}(\lambda)$ are linearly independent in $\mathbb{C}[S_n]$ and, under left multiplication, span a copy of the irrep~$\mathscr{V}_\lambda$. They are idempotent,
\begin{equation}
Y_T^2=Y_T,
\end{equation}
but not orthogonal projectors: generally,
\begin{equation}
Y_T Y_{T'} \neq 0 \qquad \text{for } T \neq T'.
\end{equation}

To obtain orthogonal projectors, one uses \emph{generalized Young symmetrizers} (GYS), also known as Hermitian Young operators~\cite{keppeler2014hermitian}. They are defined recursively:
\begin{equation}
\mathcal{Y}_T \;=\; \left( \mathcal{Y}_{\mathsf{pre}(T)} \otimes \openone \right) Y_T \left( \mathcal{Y}_{\mathsf{pre}(T)} \otimes \openone \right), 
\qquad \mathcal{Y}_T = Y_T \ \ \text{for}\ \ n \leq 2,
\label{eq:GYS}
\end{equation}
where $\mathsf{pre}(T)$ is obtained by removing the box containing the largest number from $T$.

\textit{Example.}
\begin{equation}
\mathsf{pre}\!\left(\raisebox{-1.1em}{\footnotesize\young(13,2,4)}\right)
=\raisebox{-.6em}{\footnotesize\young(13,2)}\;, 
\qquad
\mathsf{pre}\!\left(\raisebox{-1.1em}{\footnotesize\young(14,2,3)}\right)
=\raisebox{-1.1em}{\footnotesize\young(1,2,3)}\;.
\end{equation}

Here, $\mathcal{Y}_{\mathsf{pre}(T)}$ belongs to the group algebra $\mathbb{C}[S_{n-1}]$. The factor $\otimes \openone$ appends the removed number as a trivial one-cycle, embedding it into $\mathbb{C}[S_n]$ so that the whole operator acts on $n$ elements. However, since such one-cycles are usually understood implicitly once this embedding is made, we will omit $\openone$ in what follows and write simply $\mathcal{Y}_{\mathsf{pre}(T)}$.
These operators satisfy
\begin{equation}
\mathcal{Y}_T \,\mathcal{Y}_{T'} \;=\; \delta_{T,T'}\, \mathcal{Y}_T,
\label{eq:GYSorth}
\end{equation}
and thus form a complete set of mutually orthogonal idempotents in $\mathbb{C}[S_n]$. Their images under left multiplication give an explicit decomposition of $\mathbb{C}[S_n]$ into its irreducible components, each spanned by $\{\mathcal{Y}_T \,\mid\, T \in \mathrm{SYT}(\lambda)\}$ for fixed $\lambda \vdash n$. For further details and examples, see Ref.~\cite{d2021dynamical}.

\subsection{\boldmath The alternating group $A_n$}

The \emph{alternating group} $A_n$ is defined as the subgroup of the symmetric group $S_n$ consisting of all even permutations:
\begin{equation}
A_n \;:=\; \{ \sigma \in S_n \,\mid\, \mathrm{sign}(\sigma)=+1 \}.
\end{equation}
It is a normal subgroup of $S_n$ of index~2, with order $|A_n| = n!/2$, and can be equivalently characterized as the kernel of the sign homomorphism. The group $A_n$ is generated by $3$-cycles. For $n \geq 5$, it is \emph{non-solvable}, i.e., it admits no finite sequence of nested normal subgroups with abelian quotients (in fact, it is \emph{simple}, meaning it has no nontrivial normal subgroups). This has important implications. For instance, the general quintic is not solvable by radicals, since its Galois group is generically isomorphic to $S_5$, which contains the non-solvable group $A_5$ as a normal subgroup. For small $n$, one finds $A_3 \cong \mathbb{Z}_3$ (cyclic of order~3), $A_4$ of order~12 (not simple), and $A_5$ of order~60, the smallest non-Abelian simple group.

The character table of $A_n$ is obtained from that of $S_n$ by restricting characters to even permutations. In this process, certain conjugacy classes of $S_n$ split into two distinct classes in $A_n$, namely those whose elements are conjugate in~$S_n$ but not in $A_n$. Likewise, irreps of $S_n$ labeled by non-self-conjugate partitions $\lambda \neq \lambda^T$ remain irreducible when restricted to $A_n$, although they become equivalent, $D^\lambda \cong D^{\lambda^T}$, within $A_n$. In contrast, self-conjugate irreps $\lambda = \lambda^T$ split into two inequivalent irreps of $A_n$ of equal dimension, which we denote by $\lambda a$ and $\lambda b$ (this notation is nonstandard but convenient for our purposes). 

Moreover, any odd permutation $\sigma \in S_n \setminus A_n$ swaps the two corresponding irreducible subspaces:
\begin{equation}
\mathscr{K}_{\lambda a}\stackrel{\sigma}{\longleftrightarrow}\mathscr{K}_{\lambda b},\qquad \sigma \in S_n \setminus A_n.
\end{equation}

\noindent\textit{Proof.} Assume $\lambda=\lambda^T$. Recalling Eq.~(\ref{U}) we have
\begin{equation}
V^2 D^\lambda(\sigma)(V^\dagger)^2=\mathrm{sign}(\sigma)V D^\lambda(\sigma)V^\dagger=D^\lambda(\sigma)\qquad \mbox{for all $\sigma\in S_n$},
\end{equation}
so by Schur's lemma and the unitarity of $V$, it follows that $V^2= \openone_\lambda$. Define orthogonal projectors on $\mathscr{K}_\lambda$:
\begin{equation}
\mathscr{P}_a={\openone_\lambda+V\over 2},\qquad \mathscr{P}_b={\openone_\lambda-V\over 2}, 
\end{equation}
and the subspaces 
\begin{equation}
\mathscr{K}_{\lambda a}:=\mathscr{P}_a\mathscr{K}_{\lambda},\qquad \mathscr{K}_{\lambda b}:=\mathscr{P}_b\mathscr{K}_{\lambda} .
\end{equation}
We can check that these are invariant  under the action of $A_n$:
\begin{equation}
D^\lambda(\sigma)\mathscr{P}_a={1\over2}\left[D^\lambda(\sigma)+D^\lambda(\sigma) V\right]={1\over2}\left[D^\lambda(\sigma)+V D^\lambda(\sigma)\right]
=\mathscr{P}_aD^\lambda(\sigma),\qquad \mbox{for all $\sigma\in A_n$},
\end{equation}
and
\begin{equation}
D^{\lambda}(\sigma)\mathscr{K}_{\lambda a}=D^\lambda(\sigma)\mathscr{P}_a\mathscr{K}_{\lambda}=\mathscr{P}_aD^\lambda(\sigma)\mathscr{K}_{\lambda}
=\mathscr{P}_a\mathscr{K}_{\lambda}=\mathscr{K}_{\lambda a}
,\qquad \mbox{for all $\sigma\in A_n$} .
\end{equation}
The same holds for $\mathscr{P}_b$ and $\mathscr{K}_{\lambda b}$.

If instead $\sigma\in S_n \setminus A_n$ (odd), $\mathrm{sign}(\sigma)=-1$, and:
\begin{equation}
D^\lambda(\sigma)\mathscr{P}_a={1\over2}\left[D^\lambda(\sigma)+D^\lambda(\sigma) V\right]={1\over2}\left[D^\lambda(\sigma)-V D^\lambda(\sigma)\right]
=\mathscr{P}_bD^\lambda(\sigma),\qquad \mbox{for all $\sigma\in S_n\backslash A_n$},
\end{equation}
which implies
\begin{equation}
D^{\lambda}(\sigma)\mathscr{K}_{\lambda a}=D^\lambda(\sigma)\mathscr{P}_a\mathscr{K}_{\lambda}=\mathscr{P}_bD^\lambda(\sigma)\mathscr{K}_{\lambda}
=\mathscr{P}_b\mathscr{K}_{\lambda}=\mathscr{K}_{\lambda b}
,\qquad \mbox{for all $\sigma\in S_n\backslash A_n$}.
\end{equation}
This means that for $\sigma$ odd, $\mathscr{K}_{\lambda a}\stackrel{\sigma}{\longrightarrow}\mathscr{K}_{\lambda b}$. The reverse follows identically, proving that odd permutations swap the two subspaces. $\blacksquare$

All the above is illustrated in the character tables of $A_4$ and $A_5$ in Table~\ref{ct a4 & a5}, and the same structure holds for all $A_n$. For $A_4$, the classes $[2,1^2]$ and $[4]$ are absent, as they consist entirely of odd permutations. The class $[3,1]$, corresponding to $3$-cycles, splits into two since, for example, $(1,3,2)=(1,2)(1,2,3)(1,2)$ but $(1,2)\not\in A_4$. From Eq.~(\ref{conjugation}), the irreps $[4]$ and $[3,1]$ become equivalent to $[1^4]$ and $[2,1^2]$, respectively, because $\mathrm{sign}(\sigma)=1$ for all $\sigma\in A_4$; hence, only one representative from each pair is retained in the character table—specifically, the one with $\ell(\lambda)<\ell(\lambda^T)$. The irrep $[2^2]$ is self-conjugate and splits into two inequivalent $1$-dimensional irreps in $A_4$, denoted $[2^2]a$ and $[2^2]b$. Its characters in the classes $[1^4]$ and $[2^2]$ are precisely half their values in $S_4$.

An analogous analysis applies to $A_5$. Here, for example, the conjugacy class $[3,1^2]$ (the $3$-cycles) does \emph{not} split, since \mbox{$(1,3,2)=(1,2)(4,5)(1,2,3)(1,2)(4,5)$}, with $(1,2)(4,5)\in A_5$. As in $A_4$, self-conjugate irreps split \mbox{($[3,1^2] \to [3,1^2]a \oplus [3,1^2]b$)}, and their characters take (algebraically) conjugate values precisely on the split classes~$[5]a$ and $[5]b$.
\begin{table}[htb]
$$
\begin{array}{cc}
\renewcommand{\arraystretch}{1.25}
\begin{array}[t]{c|cccc}
\text{Class} & [1^4] & [2^2] & [3,1]a &[3,1]b \\
\text{Size}   & 1      & 3       & 4         & 4 \\
\hline
\chi^{[4]}     & 1 &  1 &  1 &  1 \\
\chi^{[3,1]}   & 3  & -1 &  0 &  0 \\
\chi^{[2^2]a}   & 1 &  1 & \zeta_3 &  \zeta_3^2 \\
\chi^{[2^2]b}   & 1 &  1 & \zeta_3^2 &  \zeta_3 \\
\end{array}
\hspace{4em} 
&
\hspace{4em} 
\renewcommand{\arraystretch}{1.25}
\begin{array}[t]{c|ccccc}
\text{Class} & [1^5]  & [2^2,1] & [3,1^2] & [5]a & [5]b \\
\text{Size}   &   1      &   15     &   20   &   12  & 12 \\
\hline
\chi^{[5]}     &  1 &  1 &  1 &  1 &  1 \\
\chi^{[4,1]}  &  4 &  0 &  1 & -1 & -1 \\
\chi^{[3,2]}   &  5 &  1 & -1 &  0 &  0 \\
\chi^{[3,1^2]a} &  3 &  -1 & 0 & z & z^* \\
\chi^{[3,1^2]b} &  3 &  -1 & 0 & z^* & z \\
\end{array}
\renewcommand{\arraystretch}{1}
\\
& \\[-.5em]
\fbox{$A_4$}   \hspace{4em} 
&
\hspace{4em} 
\fbox{$A_5$}
\end{array}
$$
\caption{Character tables of $A_4$ and $A_5$, where $\zeta_N=\mathrm{e}^{2\pi i/N}$ is the $N$-th root of unity, and $z=(1+\sqrt5)/2$ with $z^*=(1-\sqrt5)/2$ denoting its algebraic conjugate.}
\label{ct a4 & a5}
\end{table}

These tables, and other group-theoretic data for $A_n$, can be computed using \emph{Sage}~\cite{sage} or \emph{Magma}~\cite{magma}, both of which offer free online interfaces.

This structure is essential for our construction: it specifies which invariant subspaces remain irreducible under~$A_n$, how they intertwine with the action of $S_n$, and thereby determines which subspaces can be used for perfect parity identification, as explained in the main text.

\section{\boldmath The parity-detecting state $|\psi_e\rangle$ in the computational basis}
 
In the main text, we showed that two distinct families of states $|\psi_e\rangle$ enable parity identification with certainty whenever $d \ge d_{\rm min}$. These were specified in the Schur--Weyl basis, which makes the decomposition in Eq.~(\ref{SW-supp}) explicit. Here we derive their expression directly in the computational basis, without computing the full change-of-basis matrix.

\subsection{\boldmath Three qubits}

The Schur-Weyl decomposition of $(\mathbb{C}^2)^{\otimes3}$, Eq.~(\ref{SW-supp}), in symbolic form is
\begin{equation}
\begin{array}{rccccc}
&\ydiagram{1}\,{}^{\otimes 3}&=&\ydiagram{3}&\oplus&\ydiagram{2,1}\\[1em]
d_\lambda:&&&1&&2\\
m_\lambda:&&&4&&2\\
\text{dim. balance:}&2^3&=&1\times 4&+&2\times 2
\end{array}
\end{equation}
where $d_\lambda$ and $m_\lambda$ are the dimensions of the $S_3$ irreps and their multiplicities (i.e., dimensions of the $\SU(2)$ irreps), verifying the total dimension balance.

The self-conjugate irrep $[2,1]$ appears in this decomposition, so we project onto one of the two $A_3$-invariant subspaces, $\mathscr{K}_{[2,1]a}\otimes\mathscr{H}_{[2,1]}$ or $\mathscr{K}_{[2,1]b}\otimes\mathscr{H}_{[2,1]}$, using Eq.~(\ref{proj gen}). For instance,
\begin{equation}
\mathscr{P}_{[2,1]b}=\frac{d_{[2,1]}/2}{3!/2}\sum_{\sigma\in A_3}\overline{\chi_{[2,1]b}(\sigma)}\,\sigma,\qquad \sigma\in \mathrm{End}\big((\mathbb{C}^2)^{\otimes 3}\big),
\label{proj}
\end{equation}
where projecting onto the other subspace, labeled $[2,1]a$, would yield an equally valid parity-detecting state.

\begin{table}[htb]
$$
\begin{array}{cc}
\renewcommand{\arraystretch}{1.25}
\begin{array}[t]{c|ccc}
\text{Class} & [1^3] & [2,1] & [3] \\
\text{Size}  & 1 & 3 & 2 \\
\hline
\chi^{[3]}    & 1 &  1 &  1 \\
\chi^{[1^3]}  & 1 & -1 &  1 \\
\chi^{[2,1]}  & 2 &  0 & -1 \\
\end{array}
\hspace{4em}
&
\hspace{4em}
\renewcommand{\arraystretch}{1.25}
\begin{array}[t]{c|ccc}
\text{Class} & [1^3]  & [3]a & [3]b \\
\text{Size}  & 1      & 1    & 1    \\
\hline
\chi^{[3]}      & 1 & 1 & 1 \\
\chi^{[2,1]a}   & 1 & \zeta_3 & \zeta_3^2 \\
\chi^{[2,1]b}   & 1 & \zeta_3^2 & \zeta_3 \\
\end{array}
\renewcommand{\arraystretch}{1}
\\[.2em]
\fbox{$S_3$} \hspace{4em} & \hspace{4em} \fbox{$A_3$}
\end{array}
$$
\caption{Character tables of $S_3$ and $A_3$, where $\zeta_3=\mathrm{e}^{2\pi i/3}$.}
\label{ct s3 & a3}
\end{table}

The conjugacy classes of $A_3$ are:
\begin{equation}
C_{[1^3]}=\{e\},\qquad C_{[3]a}=\{(1,2,3)\},\qquad C_{[3]b}=\{(1,3,2)\}.
\end{equation}
Thus,
\begin{equation}
\mathscr{P}_{[2,1]b}=\frac{1}{3}\big[e+\zeta_3(1,2,3)+\zeta_3^2(1,3,2)\big],
\end{equation}
and one checks its idempotency: $\mathscr{P}_{[2,1]b}^2=\mathscr{P}_{[2,1]b}$.

Parity-detecting states are then obtained as $|\psi_e\rangle=\mathscr{P}_{[2,1]b}|\psi_0\rangle$, for {\em any} $|\psi_0\rangle$ not annihilated by $\mathscr{P}_{[2,1]b}$. Choosing $|\psi_0\rangle=|011\rangle$ yields (up to normalization):
\begin{equation}
|\psi_e\rangle = e|011\rangle+\zeta_3(1,2,3)|011\rangle+\zeta_3^2(1,3,2)|011\rangle
=|011\rangle+\zeta_3|101\rangle+\zeta_3^2|110\rangle,
\end{equation}
which is Eq.~(\ref{psie for n=3}) in the main text. This choice of $|\psi_0\rangle$ is natural since it corresponds to a valid SSYT of shape $[2,1]$:
$$
\raisebox{-1.1em}{\footnotesize\young(00,1)} \qquad
\raisebox{-1.1em}{\footnotesize\young(01,1)} 
$$

\subsection{Four qubits}

For $(\mathbb{C}^2)^{\otimes 4}$ we have
\begin{equation}
\begin{array}{rccccccc}
&\ydiagram{1}\,{}^{\otimes 4}&=&\ydiagram{4}&\oplus&\ydiagram{3,1}&\oplus&\ydiagram{2,2}\\[1em]
d_\lambda:&&&1&&3&&2\\
m_\lambda:&&&5&&3&&1\\
\text{dim. balance:}&2^4&=&1\times 5&+&3\times 3&+&2\times1
\end{array}
\end{equation}
The conjugacy classes (see Table~\ref{ct a4 & a5}) are
\begin{align}
C_{[1^4]}&=\{e\},\\
C_{[2^2]}&=\{(1,2)(3,4),(1,3)(2,4),(1,4)(2,3)\},\\
C_{[3,1]a}&=\{(1,2,3),(1,4,2),(1,3,4),(2,4,3)\},\\
C_{[3,1]b}&=\{(1,3,2),(1,2,4),(1,4,3),(2,3,4)\}.
\end{align}
The Schur-Weyl decomposition contains the self-conjugate irrep $[2^2]$. The corresponding projector is
\begin{multline}
\mathscr{P}_{[2^2]a}\propto e+(1,2)(3,4)+(1,3)(2,4)+(1,4)(2,3)+\\
\zeta_3\left[(1,3,2)+(1,2,4)+(1,4,3)+(2,3,4)\right]+\\
\zeta_3^2\left[(1,2,3)+(1,4,2)+(1,3,4)+(2,4,3)\right].
\end{multline}
The only SSYD of shape $[2^2]$ is
$$
\raisebox{-.6em}{\footnotesize\young(00,11)}\;,
$$
hence the choice $|\psi_0\rangle=|0011\rangle$. We obtain the parity-detecting state
\begin{equation}
|\psi_e\rangle = |0011\rangle + |1100\rangle 
+ \zeta_3 \left(|0101\rangle + |1010\rangle\right)
+ \zeta_3^2 \left(|0110\rangle + |1001\rangle\right),
\end{equation}
which is Eq.~(\ref{M4}) in the main text.

\subsection{Five qutrits}

For $(\mathbb{C}^3)^{\otimes 5}$ we have
\begin{equation}
\begin{array}{rccccccccccc}
&\ydiagram{1}\,{}^{\otimes 5}&=&\ydiagram{5}&\oplus&\ydiagram{4,1}&\oplus&\ydiagram{3,2}&\oplus&\ydiagram{3,1,1}&\oplus&\ydiagram{2,2,1}\\[2em]
d_\lambda:&&&1&&4&&5&&6&&5\\
m_\lambda:&&&21&&24&&15&&6&&3\\
\text{dim. balance:}&3^5\,(243)&=&1\times 21\,(21)&+&4\times 24\, (96)&+&5\times15\,(75)&+&6\times6\,(36)&+&5\times 3\,(15)
\end{array}
\end{equation}

The conjugacy classes (see Table~\ref{ct a4 & a5}) are
\begin{align}
C_{[1^5]}&=\{e\},\\[.7em]
C_{[2^2,1]}&=\{
(1,2)(3,4),\,
(1,2)(3,5),\,
(1,3)(2,4),\,
(1,3)(2,5),\,
(1,4)(2,3),\nonumber\\
&\kern5em(1,4)(2,5),\,
(1,5)(2,3),\,
(1,5)(2,4),\,
(2,3)(4,5),\,
(2,4)(3,5),\nonumber\\
&\kern10em(2,5)(3,4),\,
(3,4)(2,5),\,
(3,4)(1,5),\,
(3,5)(1,4),\,
(4,5)(1,3)
\},\\[.7em]
C_{[3,1^2]}&=\{(1,2,3),\,
(1,2,4),\,
(1,2,5),\,
(1,3,2),\,
(1,3,4),\nonumber\\
&\kern5em(1,3,5),\,
(1,4,2),\,
(1,4,3),\,
(1,4,5),\,
(1,5,2),\nonumber\\
&\kern10em(1,5,3),\,
(1,5,4),\,
(2,3,4),\,
(2,3,5),\,
(2,4,3),\nonumber\\
&\kern15em(2,4,5),\,
(2,5,3),\,
(2,5,4),\,
(3,4,5),\,
(3,5,4)\},\\[.7em]
C_{[5]a}&=\{(1,2,3,5,4),\,
(1,2,4,3,5),\,
(1,2,5,4,3),\,
(1,3,4,5,2),\,
(1,3,5,2,4),\,
(1,3,2,5,4),\nonumber\\
&\kern5em(1,4,5,2,3),\,
(1,4,2,3,5),\,
(1,4,3,2,5),\,
(1,5,3,2,4),\,
(1,5,4,3,2),\,
(1,5,2,4,3)\},\\[.7em]
C_{[5]b}&=\{(1,2,3,4,5),\,
(1,2,4,5,3),\,
(1,2,5,3,4),\,
(1,3,4,2,5),\,
(1,3,5,4,2),\,
(1,3,2,5,4),\nonumber\\
&\kern5em(1,4,5,3,2),\,
(1,4,2,3,5),\,
(1,4,3,5,2),\,
(1,5,4,2,3),\,
(1,5,2,4,3),\,
(1,5,3,2,4)
\}.
\end{align}
With this information and the character table of $A_5$ (Table~\ref{ct a4 & a5}) one can compute $\mathscr{P}_{[3,1^2]a}$.

We note that there is only one SSYT of shape $[3,1^2]$ with content $00012$:
$$
\raisebox{-1.2em}{\footnotesize\young(000,1,2)}\;,
$$
Hence, $\mathscr{P}_{[3,1^2]a}|00012\rangle$ gives the desired parity-detecting state. The result is in Eq.~(\ref{self_5}) of the main text.

\subsection{Four qutrits}

For $(\mathbb{C}^3)^{\otimes 4}$ we have
\begin{equation}
\begin{array}{rccccccccc}
&\ydiagram{1}\,{}^{\otimes 4}&=&\ydiagram{4}&\oplus&\ydiagram{3,1}&\oplus&\ydiagram{2,2}&\oplus&\ydiagram{2,1,1}\\[2em]
d_\lambda:&&&1&&3&&2&&3\\
m_\lambda:&&&15&&15&&6&&3\\
\text{dim. balance:}&3^4&=&1\times15&+&3\times15&+&2\times6&+&3\times3
\end{array}
\end{equation}

We now illustrate how to construct a parity-detecting state using conjugate pairs. Here, the relevant irreps are $[3,1]$ and $[2,1^2]$ (second and fourth on the right-hand side). We employ the generalized Young symmetrizers introduced earlier.

\subsubsection{Irrep $[3,1]$}

For $\lambda=[3,1]$, with YD $\ydiagram{3,1}$\,, the SYT are:
\begin{equation}
\mathrm{SYT}([3,1]) \;=\;
\left\{\, 
\raisebox{-.6em}{\footnotesize\young(123,4)}, \;
\raisebox{-.6em}{\footnotesize\young(124,3)}, \;
\raisebox{-.6em}{\footnotesize\young(134,2)} 
\,\right\}.
\end{equation}

To obtain a basis for this irrep, we compute the corresponding GYSs $\mathcal{Y}_T$ and apply them to 4-particle states with three $0$-excitations and one $1$-excitation. There is only one SSYT compatible with this content:
\begin{equation*}
\raisebox{-.6em}{\footnotesize\young(000,1)}\;.
\end{equation*}
Thus, such states have support only on a single copy of the irrep $[3,1]$, which greatly simplifies the construction.
From Eq.~(\ref{eq:GYS}), it follows that
\begin{align}
\mathcal{Y}_{\raisebox{-1em}{\tiny\young(123,4)}}\,&=
\mathcal{Y}_{\raisebox{-.4em}{\tiny\young(123)}}\, Y_{\raisebox{-1em}{\tiny\young(123,4)}}\, \mathcal{Y}_{\raisebox{-.4em}{\tiny\young(123)}}=
Y_{\raisebox{-.4em}{\tiny\young(12)}}\,Y_{\raisebox{-.4em}{\tiny\young(123)}}\, Y_{\raisebox{-.4em}{\tiny\young(12)}}
\, Y_{\raisebox{-1em}{\tiny\young(123,4)}}\, 
Y_{\raisebox{-.4em}{\tiny\young(12)}}\,Y_{\raisebox{-.4em}{\tiny\young(123)}}\, Y_{\raisebox{-.4em}{\tiny\young(12)}}\;,
\label{GYS 1}\\[0.5em]
\mathcal{Y}_{\raisebox{-1em}{\tiny\young(124,3)}}\,&=
\mathcal{Y}_{\raisebox{-1em}{\tiny\young(12,3)}}\, Y_{\raisebox{-1em}{\tiny\young(124,3)}}\, \mathcal{Y}_{\raisebox{-1em}{\tiny\young(12,3)}}=
Y_{\raisebox{-.4em}{\tiny\young(12)}}\,Y_{\raisebox{-1em}{\tiny\young(12,3)}}\, Y_{\raisebox{-.4em}{\tiny\young(12)}}
\, Y_{\raisebox{-1em}{\tiny\young(124,3)}}\, 
Y_{\raisebox{-.4em}{\tiny\young(12)}}\,Y_{\raisebox{-1em}{\tiny\young(12,3)}}\, Y_{\raisebox{-.4em}{\tiny\young(12)}}\;,
\label{GYS 2}\\[0.5em]
\mathcal{Y}_{\raisebox{-1em}{\tiny\young(134,2)}}\,&=
\mathcal{Y}_{\raisebox{-1em}{\tiny\young(13,2)}}\, Y_{\raisebox{-1em}{\tiny\young(134,2)}}\, \mathcal{Y}_{\raisebox{-1em}{\tiny\young(13,2)}}=
Y_{\raisebox{-1em}{\tiny\young(1,2)}}\,Y_{\raisebox{-1em}{\tiny\young(13,2)}}\, Y_{\raisebox{-1em}{\tiny\young(1,2)}}
\, Y_{\raisebox{-1em}{\tiny\young(134,2)}}\, 
Y_{\raisebox{-1em}{\tiny\young(1,2)}}\,Y_{\raisebox{-1em}{\tiny\young(13,2)}}\, Y_{\raisebox{-1em}{\tiny\young(1,2)}}\;.
\label{GYS 3}\\\nonumber
\end{align}
%where $\openone$ in $\mathcal{Y}_{\mathsf{pre}(T)}\otimes\openone$ is understood in order not to clam the expressions.

Using Eq.~(\ref{eq:YT}), one can readily compute the standard Young symmetrizers appearing on the right-hand side of Eqs.~(\ref{GYS 1})--(\ref{GYS 3}). For example:
\begin{align}
Y_{\raisebox{-.4em}{\tiny\young(12)}}\,&={1\over2}\left[e+(1,2)\right],\\
Y_{\raisebox{-.4em}{\tiny\young(123)}}\,&={1\over6}\left[e+(1,2)+(1,3)+(2,3)+(1,2,3)+(1,3,2)\right],\\
Y_{\raisebox{-1em}{\tiny\young(123,4)}}\,&={1\over 8}\left[e+(1,2)+(1,3)+(2,3)+(1,2,3)+(1,3,2)-\right.\nonumber\\
&\kern10em \left.(1,4)-(1,4,2)-(1,4,3)-(1,4)(2,3)-(1,4,2,3)-(1,4,3,2)\right].
\end{align}
Here we also note some useful simplifications, e.g.,
\begin{equation}
\mathcal{Y}_{\raisebox{-1em}{\tiny\young(123,4)}}\,=
Y_{\raisebox{-.4em}{\tiny\young(123)}}
\, Y_{\raisebox{-1em}{\tiny\young(123,4)}}\, 
Y_{\raisebox{-.4em}{\tiny\young(123)}}\,=
Y_{\raisebox{-1em}{\tiny\young(123,4)}}\, 
Y_{\raisebox{-.4em}{\tiny\young(123)}}\;.
\end{equation}
Thus,
\begin{equation}
\mathcal{Y}_{\raisebox{-1em}{\tiny\young(123,4)}}\,|0001\rangle=Y_{\raisebox{-1em}{\tiny\young(123,4)}}\, 
Y_{\raisebox{-.4em}{\tiny\young(123)}}\;|0001\rangle
=Y_{\raisebox{-1em}{\tiny\young(123,4)}}\, 
|0001\rangle=-{|1000\rangle+|0100\rangle+|0010\rangle-3|0001\rangle\over 4}.
\end{equation}
Normalizing this state yields the first basis element of irrep $[3,1]$:
\begin{equation}
|u^{[3,1]}_1\rangle={|1000\rangle+|0100\rangle+|0010\rangle-3|0001\rangle\over 2\sqrt 3}.
\end{equation}

Proceeding similarly with $\mathcal{Y}_{\raisebox{-1em}{\tiny\young(124,3)}}\,|0010\rangle$ and $\mathcal{Y}_{\raisebox{-1em}{\tiny\young(134,2)}}\,|0100\rangle$, we obtain the remaining two basis vectors:
\begin{align}
|u^{[3,1]}_2\rangle&= {|1000\rangle+|0100\rangle-2|0010\rangle\over\sqrt6},\\[.0em]
|u^{[3,1]}_3\rangle&={|1000\rangle-|0100\rangle\over\sqrt2}.
\end{align}
These vectors, together with $|u^{[3,1]}_1\rangle$, form an orthonormal basis $\{|u^{[3,1]}_k\rangle\}_{k=1}^3$ that carries the irrep $[3,1]$ of $S_4$, namely:
\begin{equation}
\sigma |u^{[3,1]}_k\rangle =\sum_{l=1}^3 \big[D^{[3,1]}(\sigma)\big]_{l,k} |u^{[3,1]}_l\rangle.
\end{equation}
The representation matrices $D^{[3,1]}(\sigma)$ coincide with those obtained from Yamanouchi's construction~\cite{audenaert2006digest,walther2024}. Explicitly, for the adjacent transpositions (which generate $S_4$) one has:
\begin{equation}
D^{[3,1]}\big((12)\big)=\begin{pmatrix} 1 &0&0\\[.2em] 0&1&0\\[.2em]0&0&-1\end{pmatrix},\quad
D^{[3,1]}\big((23)\big)=\begin{pmatrix} 1 &0&0\\[.2em] 0&-{1\over2}&{\sqrt3\over2}\\[.2em]0&{\sqrt3\over2}&{1\over2}\end{pmatrix},\quad
D^{[3,1]}\big((34)\big)=\begin{pmatrix} -{1\over3} &{2\sqrt2\over3}&0\\[.2em] {2\sqrt2\over3}&{1\over3}&0\\[.2em]0&0&1\end{pmatrix}.
\label{irrep [3,1]}
\end{equation}
This completes the explicit construction of the $[3,1]$ irrep basis.

\subsubsection{Irrep $[2,1^2]$}

For the conjugate irrep $\lambda=[2,1^2]$, with YD $\ydiagram{2,1,1}$\,, the SYTs are:
\begin{equation}
\mathrm{SYT}([2,1^2]) \;=\;
\left\{\, 
\raisebox{-1.1em}{\footnotesize\young(14,2,3)}, \;
\raisebox{-1.1em}{\footnotesize\young(13,2,4)}, \;
\raisebox{-1.1em}{\footnotesize\young(12,3,4)} 
\,\right\}.
\end{equation}

We then compute the corresponding GYSs and apply them to $4$-particle states containing two $0$-excitations, one $1$-excitation, and one $2$-excitation, since such states have support in a single copy of the irrep $[2,1^2]$ (only one SSYT is possible with this content). After some work, we obtain the following basis:
\begin{align}
|u^{[2,1^2]}_1\rangle&= 
{
2 |0012\rangle 
- 2 |0021\rangle 
- |0102\rangle 
+ |0120\rangle 
+ |0201\rangle 
- |0210\rangle 
- |1002\rangle 
+ |1020\rangle 
+ |2001\rangle 
- |2010\rangle
\over 4},
\\[.5em]
|u^{[2,1^2]}_2\rangle&={3 |0102\rangle 
- |0120\rangle 
- 3 |0201\rangle 
+ |0210\rangle 
- 3 |1002\rangle 
+ |1020\rangle 
+ 2 |1200\rangle 
+ 3 |2001\rangle 
- |2010\rangle 
- 2 |2100\rangle\over 4\sqrt3}, \\[.5em]
|u^{[2,1^2]}_3\rangle&= {|0120\rangle 
- |0210\rangle 
- |1020\rangle 
+ |1200\rangle 
+ |2010\rangle 
- |2100\rangle
\over\sqrt6} .
\end{align}
This orthonormal basis carries the irrep $[2,1^2]$. The corresponding representation matrices coincide with those obtained via Yamanouchi's construction:
\begin{equation}
D^{[2,1^2]}\big((12)\big)=\begin{pmatrix} 1 &0&0\\[.2em] 0&-1&0\\[.2em]0&0&-1\end{pmatrix},\quad
D^{[2,1^2]}\big((23)\big)=\begin{pmatrix} -{1\over2} &{\sqrt3\over2}&0\\[.2em] {\sqrt3\over2}&{1\over2}&0\\[.2em]0&0&-1\end{pmatrix},\quad
D^{[2,1^2]}\big((34)\big)=\begin{pmatrix}-1&0&0 \\[.2em] 0& -{1\over3} &{2\sqrt2\over3} \\[.2em] 0 &{2\sqrt2\over3}&{1\over3}\end{pmatrix}.
\label{irrep [2,1^2]}
\end{equation}

\subsubsection{The parity-detecting state $|\psi_e\rangle$}

As explained in previous sections, since $[3,1]$ and $[2,1^2]$ form a conjugate pair of irreps, one has $D^{[2,1^2]}\cong \mathrm{sign}\otimes D^{[3,1]}$. Hence, there exists a change-of-basis matrix $T$ such that
\begin{equation}
D^{[2,1^2]}(\sigma)= \mathrm{sign}(\sigma)\; T D^{[3,1]}(\sigma) T^{\dagger}, \qquad \mbox{for all $\sigma\in S_4$}.
\end{equation}
Examining Eqs.~(\ref{irrep [3,1]}) and~(\ref{irrep [2,1^2]}) shows that
\begin{equation}
T=\begin{pmatrix}0&0&1\\ 0&-1&0\\ 1&0&0 \end{pmatrix}.
\end{equation}
Following our notation in the main text, we relabel the two bases as
\begin{align}
|v^{[3,1]}_k\rangle|s_+\rangle&:=|u^{[3,1]}_k\rangle,\qquad \mbox{for $k=1,2,3$},\\
|v^{[3,1]}_1\rangle|s_-\rangle&:=|u^{[2,1^2]}_3\rangle,\\
|v^{[3,1]}_2\rangle|s_-\rangle&:=-|u^{[2,1^2]}_2\rangle,\\
|v^{[3,1]}_3\rangle|s_-\rangle&:=|u^{[2,1^2]}_1\rangle.
\end{align}
In this basis, the irreps $[3,1]$ and $[2,1^2]$ become identical when restricted to $A_4$, with $|s_\pm\rangle$ serving as the multiplicity basis. Consequently,
\begin{align}
\sigma_{\rm even}\, |v^{[3,1]}_k\rangle|s_\pm\rangle&=\sum_{l=1}^3 \big[D^{[3,1]}(\sigma_{\rm even})\big]_{l,k}\,|v^{[3,1]}_l\rangle|s_\pm\rangle,\\
\sigma_{\rm odd}\, |v^{[3,1]}_k\rangle|s_\pm\rangle&=\pm\sum_{l=1}^3 \big[D^{[3,1]}(\sigma_{\rm odd})\big]_{l,k}\,|v^{[3,1]}_l\rangle|s_\pm\rangle,
\end{align}
for any even or odd permutation $\sigma$. Hence, any state of the form
\begin{equation}
|\psi_e\rangle= \sum_{k=1}^3 \phi_k\,|v^{[3,1]}_k\rangle\big(|s_+\rangle+|s_-\rangle\big)
\end{equation}
satisfies
\begin{equation}
\sigma_{\rm even}|\psi_e\rangle\quad \perp \quad \sigma_{\rm odd}|\psi_e\rangle,
\end{equation}
thus enabling parity identification with certainty. In particular, choosing $|\psi_e\rangle=|v^{[3,1]}_1\rangle(|s_+\rangle+|s_-\rangle)/\sqrt2$ yields the state given in Eq.~(\ref{conj_pairs_4}) of the main text.

\section{How much entangled is a parity-detecting state?}  

Here we provide further details on how we compute numerical lower bounds on the entanglement required for parity detection for $n=3$, $4$, and $5$.  

We quantify entanglement using the geometric measure of entanglement (GME)~\mbox{\cite{Wei2003geometric,Lisa2025quantifyin}}, which for a pure multipartite state $|\psi\rangle$ is defined as
\begin{equation}
E(\ket{\psi}) := 1 - \max_{\ket{\phi} \in \mathrm{SEP}} |\langle \phi | \psi \rangle|^2,
\end{equation}
where the maximization is over the set $\mathrm{SEP}$ of all separable states, \mbox{$\ket{\phi} = \bigotimes_{i=1}^n \ket{\phi_i}$}.

%to find a lower bound on the entanglement $E$

Our goal is to evaluate the GME of the least entangled state within the subspace $\bm{\mathscr{K}}_{\lambda a}:=\mathscr{K}_{\lambda a} \otimes \mathscr{H}_{\lambda}$ that enables parity discrimination, with associated projector $\mathscr{P}_{\lambda a}$. Namely, the quantity
\begin{equation}\label{eq:minmax}
    E_{\lambda a} := \min_{\ket{\psi}\in \bm{\mathscr{K}}_{\lambda a}} E(|\psi\rangle) 
    = 1 - \max_{\ket{\psi}\in  \bm{\mathscr{K}}_{\lambda a}} \max_{\ket{\phi}\in\mathrm{SEP}}|\langle\phi|\psi\rangle|^2\,.
\end{equation}
We first show that this quantity can be expressed as
\begin{equation}
    E_{\lambda a} = 1 - \max_{\ket{\phi} \in \mathrm{SEP}} \bra{\phi} \mathscr{P}_{\lambda a} \ket{\phi}.
\label{eq81}
\end{equation}

%To find lower bounds for $ E_{\lambda a}$, we need to perform two optimizations: over parity-detecting states $\ket{\psi}\in \bm{\mathscr{K}}_{\lambda a}$ and over separable states $\phi\in\text{SEP}$. We manipulate Eq.~\eqref{eq:minmax} in two steps.

\noindent\textit{Proof.}
We note that for any $\ket{\psi}\in\bm{\mathscr{K}}_{\lambda a}$
\begin{equation}\label{eq:Minmax<E}
   % \max_{\ket{\psi}\in\bm{\mathscr{K}}_{\lambda a}} 
   \max_{|\phi\rangle\in\text{SEP}}|\langle\phi|\psi\rangle|^2 
    \leq \max_{\ket{\phi} \in \mathrm{SEP}} \bra{\phi} \mathscr{P}_{\lambda a} \ket{\phi}.
\end{equation}
To see this, let $\ket{\phi^*_\psi},\ket{\phi^*}\in\text{SEP}$ be the states that maximize the overlap on the left-hand side and the expectation value on right-hand side, respectively. 
Since $\mathscr{P}_{\lambda a}$ can be written as a uniform convex combination of any orthogonal basis of $\bm{\mathscr{K}}_{\lambda a}$, in particular one including $\ket{\psi}$, the operator $\mathscr{P}_{\lambda a}-|\psi\rangle\langle\psi|$ is positive semidefinite. Thus,
\begin{equation}
  \max_{|\phi\rangle\in\text{SEP}}|\langle\phi|\psi\rangle|^2 
    =
|\langle\phi^*_\psi|\psi\rangle|^2\leq \bra{\phi^*_\psi}\mathscr{P}_{\lambda a}\ket{\phi^*_\psi}  \leq \bra{\phi^*} \mathscr{P}_{\lambda a} \ket{\phi^*}=\max_{\ket{\phi} \in \mathrm{SEP}} \bra{\phi} \mathscr{P}_{\lambda a} \ket{\phi}
,\quad \mbox{for any $|\psi\rangle\in \bm{\mathscr{K}}_{\lambda a}$}.
\label{sat}
\end{equation}
This inequality is attainable since one can construct a parity-detecting state as
\begin{equation}
    \ket{\psi}=\frac{\mathscr{P}_{\lambda a}\ket{\phi^*}}{\|\mathscr{P}_{\lambda a}\ket{\phi^*}\|} \in  \bm{\mathscr{K}}_{\lambda a}\,,
    \label{psi opt}
\end{equation}
and choosing $|\phi^*_{\psi}\rangle=|\phi^*\rangle$, we obtain
\begin{equation}
|\langle\phi^*_{\psi}|\psi\rangle|^2
= \left| \frac{\bra{\phi^*}\mathscr{P}_{\lambda a}\ket{\phi^*}}{\|\mathscr{P}_{\lambda a}\ket{\phi^*}\|} \right|^2
= \frac{\bra{\phi^*}\mathscr{P}_{\lambda a}^2\ket{\phi^*}^2}{\|\mathscr{P}_{\lambda a}\ket{\phi^*}\|^2}
= \|\mathscr{P}_{\lambda a}\ket{\phi^*}\|^2
= \bra{\phi^*} \mathscr{P}_{\lambda a} \ket{\phi^*},
\end{equation}
thus saturating Eq.~(\ref{sat}), and in turn showing that the parity-detecting state in Eq.~(\ref{psi opt}) achieves the maximum possible value of 
$\max_{|\phi\rangle\in\text{SEP}}|\langle\phi|\psi\rangle|^2$ over $\bm{\mathscr{K}}_{\lambda a}$, i.e.,
\begin{equation}
  \max_{\ket{\psi}\in  \bm{\mathscr{K}}_{\lambda a}}  \max_{|\phi\rangle\in\text{SEP}}|\langle\phi|\psi\rangle|^2  
  = \max_{\ket{\phi} \in \mathrm{SEP}} \bra{\phi} \mathscr{P}_{\lambda a} \ket{\phi},
\end{equation}
which leads to Eq.~(\ref{eq81})~$\blacksquare$

Evaluating $E_{\lambda a}$ exactly is challenging, as the brute-force optimization of Eq~(\ref{eq81})  is highly nonlinear and yields results with no guarantee of global maxima.  
Instead, we derive a rigorous lower bound using semidefinite programming (SDP), as detailed next.

\noindent\textit{SDP relaxation.}  
We outer-approximate the set SEP using SDP with symmetric extensions (DPS$_k$)~\cite{DPS2004,NavascuesDPS2009}. In this hierarchy, SEP is approximated at level $k=0$ by the set of PPT states $\varrho$, while higher levels impose both PPT and symmetry constraints on an extended system.

Exploiting the invariance of $\mathscr{P}_{\lambda a}$ under diagonal (local) unitaries,  
$U^{\otimes n}\,\mathscr{P}_{\lambda a}\,{U^\dagger}^{\otimes n}=\mathscr{P}_{\lambda a}$,  
and noting that the maximization in Eq.~(\ref{eq81}) is over the set of product states, we can fix the state of one subsystem:
\begin{equation}
    \max_{|\phi_1\cdots\phi_n\rangle}
    \bra{\phi_1\cdots\phi_n}\mathscr{P}_{\lambda a}\ket{\phi_1\cdots\phi_n} 
    = 
    \max_{|\phi_2\cdots\phi_n\rangle}
    \bra{0\phi_2\cdots\phi_n}\mathscr{P}_{\lambda a}\ket{0\phi_2\cdots\phi_n}.
\end{equation}
This leads to the computable bound:
\begin{equation}\label{eq:0PPT}
    E_{\lambda a}\geq 1-\max_{\varrho\in \mathrm{DPS}_k}\Tr\Big (\Tr_1(\ket{0}\bra{0}\otimes\openone^{\otimes (n-1)}\,\mathscr{P}_{\lambda a})\,\varrho\Big ).
\end{equation}
For $n=3$ (three qubits), $\varrho$ is a two-qubit state; since for two qubits PPT implies separability, the resulting value, $E_{[2,1] a}=5/9$, is exact.

So far, we have lower bounded the entanglement of pure parity-detecting states $|\psi_e\rangle$. To extend this to mixed states~$\rho_e$ (supported entirely on $\bm{\mathscr{K}}_{\lambda a}$), we follow Refs.~\cite{Wei2003geometric,das2016generalized} and define the GME of a generic multipartite density matrix $\rho$ via the convex roof:
\begin{equation}
E(\rho):=\min_{\{p_i,|\psi_i\rangle\}} \sum_i p_i \, E(|\psi_i\rangle),
\label{GME mixed}
\end{equation}
where the minimization is over all pure-state decompositions $\rho=\sum_i p_i |\psi_i \rangle\langle\psi_i|$.

In our case, every decomposition of $\rho_e$ lies entirely within $\bm{\mathscr{K}}_{\lambda a}$. If $\{ p^*_i,|\psi^*_i\rangle \}$ attain the minimum in Eq.~(\ref{GME mixed}), and $|\psi_0^*\rangle$ is the state with minimum GME in this optimal decomposition, then
\begin{equation}
E(\rho_e)= \sum_i p^*_i E(|\psi^*_i\rangle)
\ge \Bigg(\sum_i p^*_i \Bigg) E(|\psi^*_0\rangle)
= E(|\psi^*_0\rangle)
\ge \min_{|\psi\rangle\in \bm{\mathscr{K}}_{\lambda a}} E(|\psi\rangle)=E_{\lambda a}.
\end{equation}
Thus, the lower bounds derived in the main text, as explained above, also apply to any parity-detecting state, whether pure or mixed.

\end{document}